%% file: CMPfinal.tex
\documentclass[a4paper,10pt]{article}
\usepackage{graphicx}
\usepackage{color}\usepackage{amsmath, amsthm, slashed}
\usepackage{amssymb}
\usepackage{rotating}
\usepackage{epsfig}
\usepackage{verbatim}
\usepackage{rotating}
\usepackage{graphicx}
\newtheorem{definition}{Definition}[section]
\newtheorem{theorem}{Theorem}[section]

\newtheorem{corollary}{Corollary}[section]
\newtheorem{proposition}{Proposition}[section]
\newtheorem{lemma}{Lemma}[section]

\title{On the massive wave equation on slowly rotating Kerr-AdS spacetimes}
\author{Gustav Holzegel\thanks{Princeton University,
Department of Mathematics, Fine Hall, Washington Road,
Princeton, NJ 08544 United States}}

\begin{document}
\maketitle
\begin{abstract}
The massive wave equation $\Box_g \psi - \alpha\frac{\Lambda}{3} \psi = 0$ is studied on 
a fixed Kerr-anti de Sitter background $\left(\mathcal{M},g_{M,a,\Lambda}\right)$. 
We first prove that in the Schwarzschild case $(a=0),$ 
 $\psi$ remains uniformly bounded on the black hole exterior provided 
that $\alpha < \frac{9}{4}$, i.e.~the Breitenlohner-Freedman bound holds. Our proof is based 
on vectorfield multipliers and commutators: The total flux of the usual energy current 
arising from the timelike Killing vector field $T$ (which fails to be non-negative pointwise) is shown to be 
non-negative with the help of a Hardy inequality after integration over a spacelike slice. 
In addition to $T$, we construct a vectorfield whose energy identity captures the redshift 
producing good estimates close to the horizon. The argument is finally generalized to slowly 
rotating Kerr-AdS backgrounds. This is achieved by replacing the Killing vectorfield 
$T=\partial_t$ with $K=\partial_t + \lambda \partial_\phi$ for an appropriate $\lambda \sim a$,
which is also Killing and--in contrast to the asymptotically flat case--everywhere causal 
on the black hole exterior. The separability properties of the wave equation on Kerr-AdS 
are not used. As a consequence, the theorem also applies to spacetimes 
sufficiently close to the Kerr-AdS spacetime, as long as they admit a causal 
Killing field $K$ which is null on the horizon. 
\end{abstract}

\section{Introduction}
The study of linear wave equations on black hole spacetimes has acquired a prominent role  
within the subject of general relativity. The main reason is the expectation that understanding the mechanisms responsible for the decay of linear waves on black hole exteriors in a sufficiently robust setting provides important insights for the non-linear black hole stability problem \cite{Mihalisnotes}.

The mathematical analysis of linear waves in this context was initiated by the pioneering work of Kay and Wald establishing boundedness (up to and including the event horizon) for $\phi$ satisfying $\Box_g \phi=0$ on Schwarzschild spacetimes \cite{Wald2, KayWald}. Since then considerable progress has been achieved, especially in the last few years. Most of these recent decay and boundedness theorems for linear waves concern black hole spacetimes satisfying the vacuum Einstein equations
\begin{equation} \label{Einstein}
R_{\mu \nu} - \frac{1}{2}R g_{\mu \nu} + \Lambda g_{\mu \nu} = 0 
\end{equation}
with $\Lambda=0$ and, motivated by cosmological considerations, $\Lambda>0$. In particular, by now polynomial decay rates have been established for $\Box_g \phi=0$ on Schwarzschild \cite{DafRod2, Sterbenz, DafRod} and more recently, Kerr spacetimes \cite{DafRodKerr, Mihalisnotes, Toha2}. 
In the course of work on the decay problem, a much more robust understanding of boundedness on both Schwarzschild and Kerr spacetimes was also obtained \cite{DafRodKerr, Mihalisnotes} allowing one to prove boundedness for a large class of spacetimes, which are not exactly Schwarzschild or Kerr 
but only assumed to be sufficiently close

Many of the above results have been extended to the cosmological case, $\Lambda>0$. Here (much stronger) decay rates have been established for the wave equation on Schwarzschild-de Sitter spacetimes \cite{DafRoddS, Haefner, Melrose}. 

For further discussion we refer the reader to the lecture notes \cite{Mihalisnotes}, which among other things provide an account of previous work on these problems (sections 4.4, 5.5, 6.3 ibidem) and a comparison with results that have been obtained in the heuristic tradition (section 4.6).

In contrast to the case of a positive cosmological constant, the choice $\Lambda<0$ in (\ref{Einstein}) has remained relatively unexplored. While this problem certainly deserves mathematical attention in its own right, there is also considerable interest from high energy physics, see \cite{Gubser, Julianos}. 

In this paper, we study the equation
\begin{equation} \label{study}
\Box_g \psi - \alpha \frac{\Lambda}{3} \psi = 0
\end{equation}
on a class of spacetimes, which will include slowly rotating Kerr anti-de Sitter spacetimes \cite{CarterAdS}. These spacetimes generalize the well-known Kerr solution (the latter being the unique two-parameter family of stationary, axisymmetric asymptotically flat black hole solutions to (\ref{Einstein}) with $\Lambda=0$.). They are axisymmetric, stationary solutions of (\ref{Einstein}) with $\Lambda<0$, parametrized by their mass $M$ and angular momentum per unit mass $a=\frac{J}{M}$. Before we comment further on their geometry, let us discuss equation (\ref{study}). The main motivation to include the zeroth order term in (\ref{study}) is the case $\alpha=2$, the conformally invariant case. In pure AdS the Green's function for (\ref{study}) with $\alpha=2$ is supported purely on the light cone, which makes it a natural analogue of the equation $\Box \psi = 0$ in asymptotically flat space (and also explains the adjective ``massless'' which is sometimes used in the physics literature in connection with this choice of $\alpha$). The case $\alpha=2$ also occurs naturally in classical general relativity when studying a Maxwell field or linear gravitational perturbations in AdS \cite{Ishibashi}.

Again for the case of pure AdS, it is well known (\cite{Breitenlohner}, \cite{Bachelot}, \cite{Ishibashi}) that (\ref{study}) is only well-posed for $\alpha<\frac{5}{4}$, the so-called \emph{second Breitenlohner-Freedman bound}. While no solutions exist for $\alpha \geq \frac{9}{4}$, one has an infinite number of solutions depending on boundary conditions for $\alpha$ in the range $\frac{5}{4}\leq \alpha <\frac{9}{4}$, the latter bound being the \emph{first Breitenlohner-Freedman bound}. For general asymptotically AdS spacetimes, however, the wellposedness of (\ref{study}) has not yet been established explicitly. In this context it is essential to notice that asymptotically AdS spaces are not globally hyperbolic. To make the dynamics of (\ref{study}) well-posed suitable boundary conditions will have to be imposed on the timelike boundary of the spacetime. We will address this boundary initial value problem in detail in a seperate paper. For the purpose of the present paper, we will assume that $\alpha<\frac{9}{4}$ and that we are given a solution to (\ref{study}) which decays suitably near the AdS boundary. This pointwise ``radial'' decay depends on how close $\alpha$ is to the Breitenlohner Freedman bound and ensures in particular that there is no energy flux through the timelike boundary $\mathcal{I}$. It is precisely the decay suggested by the mode analysis of \cite{Breitenlohner} in pure AdS and expected to hold for all asymptotically AdS spacetimes. 

The global geometry of the Kerr-AdS background spacetimes is quite different from their asymptotically flat counterparts. Most notably perhaps, null-infinity is now timelike, entailing the non-globally hyperbolic nature of the spacetime mentioned above. With the boundary conditions imposed there will be no radiation flux through infinity and the only possible decay mechanism is provided by an energy flux through the horizon. Another geometric feature, which we will exploit to a great extent in the present paper, is the existence of an everywhere causal Killing vectorfield on the black hole exterior of slowly rotating Kerr-AdS spacetimes.\footnote{This is crucially different from the asymptotically flat case, where any linear combination of the two available Killing fields $K=\partial_t + \lambda \partial_\phi$ (for some constant $\lambda$) is somewhere spacelike.} This vectorfield was used previously by Hawking and Reall \cite{HawkingReall} to obtain a positive conserved energy for any matter fields satisfying the dominant energy condition. The scalar field (\ref{study}) in the subcase $\alpha \leq 0$ provides an example. In particular, the argument of \cite{HawkingReall} excluded a negative energy flux through the horizon and hence superradiance as a mechanism of instability, at least if $|a| \sqrt{-\frac{3}{\Lambda}}<r_{hoz}^2$ with $r_{hoz}$ being the location of the event horizon.

For $0<\alpha<\frac{9}{4}$, however, the energy momentum tensor associated with $\psi$ does not satisfy the dominant energy condition and \emph{the energy current associated with the causal Killing field fails to be positive pointwise}. In particular, as noted for instance in \cite{Kunduri}, the positivity argument of \cite{HawkingReall} breaks down for $0<\alpha<\frac{9}{4}$, including the most interesting case $\alpha=2$. In this paper we present a simple resolution of this problem: Using a Hardy inequality we show that the energy flux arising from the Killing field is still \emph{positive in an integrated sense}.
With the existence of the everywhere causal Killing field and its associated globally positive energy, superradiance is eliminated as an obstacle to stability.\footnote{As expected perhaps, 
the restriction on $a$ (which can be computed explicitly) becomes tighter compared to the case $\alpha \leq 0$ of Hawking and Reall.} This makes the problem much easier to deal with than the asymptotically flat case, where such a vectorfield is not available. In particular, we can avoid the intricate bootstrap and harmonic analysis techniques of \cite{DafRodKerr}, which were necessary to deal with superradiant phenomena and trapping (see also \cite{Toha2}).\footnote{Superradiance is induced by the existence of an ergosphere [i.e.~an effect which is not present in Schwarzschild] arising from the fact that the energy density associated with the Killing field $\partial_t$ can be negative inside the ergosphere. This allows for a negative energy flux through the horizon and hence an amplification of the amplitude for backscattered waves. See \cite{Wald} for a nice discussion and also \cite{Mihalisnotes} for a detailed mathematical treatment.} 

It is well-known that the notion of positive energy outlined above is not sufficient to prevent the scalar field from blowing up on the horizon in evolution. This issue was first addressed and resolved for Schwarzschild in the celebrated work of Kay and Wald \cite{Wald2, KayWald} exploiting the special symmetry properties of the background spacetime. With the recent work of Dafermos and Rodnianski, in particular their mathematical understanding of the celebrated redshift, there is now a much more stable argument available, which does not hinge on the discrete
symmetries of Schwarzschild and is in fact applicable to any black hole event horizon with positive surface gravity \cite{Mihalisnotes}. This geometric understanding of the role of the event horizon for the boundedness and decay mechanism was first developed in \cite{DafRod2} and plays a crucial role in the recent proof of boundedness \cite{DafRodKerr} and polynomial decay \cite{Mihalisnotes} of scalar waves on slowly rotating Kerr spacetimes. The boundedness statement of \cite{DafRodKerr} holds for a much more general class of spacetimes nearby Kerr, while the decay statement of \cite{Mihalisnotes} requires the background to be exactly Kerr.

For our considerations at the horizon we will adapt the ideas developed by Dafermos and Rodnianski. A vectorfield $Y$ is constructed whose energy identity, if coupled to the timelike Killing vectorfield $K$ in an appropriate way, provides control over Sobolev norms whose weights do not degenerate at the horizon. An additional complication compared to the asymptotically-flat massless case is that due to the zeroth order term in the wave equation (\ref{study}), a zeroth order flux-term of the redshift vectorfield on the horizon has the wrong sign. However, this term can, after a computation, be absorbed by the ``good'' positive terms at our disposal. With $Y$ and $K$ at hand, boundedness 
is shown adapting the argument of \cite{DafRodKerr}, taking care of the different weights in $r$ which appear due to the asymptotically hyperbolic nature of the background space.\footnote{Since angular momentum operators do not commute with the wave operator for $a\neq 0$, we have to commute the wave equation with $T$ and the redshift vectorfield $Y$ to obtain $L^2$ control of certain derivatives, which leads to control over all derivatives using elliptic estimates on the asymptotically hyperbolic spacelike slices $\Sigma$.} 

It should be emphasized that the proof does \emph{not} require the construction of a globally positive spacetime integral arising from a virial vectorfield but that it suffices to use the timelike Killing field and the redshift vectorfield alone. The fact that the more elementary statement of boundedness can be obtained from the use of these two vectorfields alone in the non-superradiant regime was observed by Dafermos and Rodnianski in \cite{DafRodKerr} in the asymptotically-flat case. Previously, even the understanding of boundedness was intriniscially tied to the understanding of a globally positive spacetime term and hence decay, see \cite{DafRod2} and also section 3.4 of \cite{Mihalisnotes}.

Finally, here is an outline of the paper. We introduce the AdS-Schwarzschild and Kerr backgrounds equipped with regular coordinate systems on the black hole exterior (defining in particular the spacelike slices we are going to work with) in section 2. In the following section the class of solutions we wish to consider is defined and the notion of vectorfield multipliers discussed. For reasons of presentation we first state and prove the boundedness theorem for Schwarzschild-AdS in section 4, before we turn to the generalization to Kerr-AdS in section 5. The paper concludes with some final remarks and future directions.

\section{The black hole backgrounds}

\subsection{Schwarzschild-AdS}
In the familiar $(t,r)$ coordinates, the Schwarzschild AdS metric reads
\begin{equation}
g = -\left(1-\frac{2M}{r}+\frac{r^2}{l^2}\right) dt^2 + \left(1-\frac{2M}{r}+\frac{r^2}{l^2}\right)^{-1} dr^2 + r^2 d\omega_2
\end{equation}
where $\Lambda = -\frac{3}{l^2}$ is the cosmological constant. This coordinate system is not well-behaved at the zeros of 
\begin{equation} \label{oneminusmu}
1-\mu = \left(1-\frac{2M}{r}+\frac{r^2}{l^2}\right)  \, .
\end{equation}
Let us define
\begin{equation}
p = \left(M l^2 + \sqrt{M^2 l^4 + \frac{l^6}{27}}\right)^{\frac{1}{3}} \textrm{ \ \ \ and \ \ \ }  q = \left(M l^2 - \sqrt{M^2 l^4 + \frac{l^6}{27}}\right)^{\frac{1}{3}} \, .
\end{equation}
Clearly, $p>0$, $q<0$ with $pq = -\frac{l^2}{3}$. The expression (\ref{oneminusmu}) has a single real root at
\begin{eqnarray} \label{rhozS}
r_{hoz} = p + q > 0 \, ,
\end{eqnarray}
the location of the black hole event horizon.
Note that for $l\rightarrow \infty$ we have $r_{hoz} \rightarrow 2M$. In general we have the estimate
\begin{eqnarray} \label{cubic}
1-\mu = \frac{r^3 + l^2 r - 2Ml^2}{r l^2} =  \frac{\left(r-r_{hoz}\right)\left(r^2 + r r_{hoz} + r_{hoz}^2 - 3pq\right)}{r l^2} \nonumber \\ \frac{\left(r-r_{hoz}\right)\left(r^2 + r r_{hoz} + r_{hoz}^2+l^2\right)}{r l^2} \geq \frac{\left(r^3-r_{hoz}^3\right)}{r l^2} \, ,
\end{eqnarray}
which will be useful later. 

Here is a coordinate system which is well behaved everywhere on the black hole exterior and the horizon:
%
%
%
%
%
%
%
%
%
%
It arises from the coordinate transformation
\begin{equation}
t^\star = t + r^\star\left(r\right) - l \arctan\left(\frac{r}{l}\right) \, ,
\end{equation}
where $r^\star\left(r\right)$ is a solution of the differential equation
\begin{equation}
\frac{dr^\star}{dr} = \frac{1}{1-\frac{2M}{r} + \frac{r^2}{l^2}} = \frac{1}{1-\mu} \textrm{ \ \ \ and \ \ \ $r^\star\left(3M\right)=0$} \, .
\end{equation}
The variable $r^\star$ is often called the tortoise coordinate. In the new $\left(t^\star, r\right)$ coordinates the metric reads
\begin{equation} \label{Schwtr}
g = -\left(1-\mu\right) \left(dt^\star\right)^2 + \frac{4M}{r\left(1+\frac{r^2}{l^2}\right)} dt^\star dr + \frac{1+ \frac{2M}{r} 
+\frac{r^2}{l^2}}{\left(1+\frac{r^2}{l^2}\right)^2} dr^2 + r^2 d\omega_2 \, ,
\end{equation}
which is clearly regular on the horizon. For notational convenience let us agree on the shorthand notation
\begin{equation}
k_{\pm} = 1 \pm \frac{2M}{r} + \frac{r^2}{l^2} \textrm{ \ \ \ and \ \ \ } k_0 = 1 + \frac{r^2}{l^2} \, .
\end{equation}
%
%
%
%
%
%
%
%
%
%
Slices $\Sigma_\tau$ of constant $t^\star$ will play a prominent role for stating energy identities in the paper. Their normal
\begin{equation}
-\nabla t^\star =  \frac{1+ \frac{2M}{r} 
+\frac{r^2}{l^2}}{\left(1+\frac{r^2}{l^2}\right)^2} \partial_{t^\star} - \frac{2M}{r \left(1+\frac{r^2}{l^2}\right)} \partial_r
\end{equation}
is everywhere timelike:
\begin{equation}
g\left(\nabla t^\star, \nabla t^\star\right) = g^{t^\star t^\star} = - \frac{k_+}{k_0^2} \, .
\end{equation}
We denote the unit-normal by
\begin{equation}
n_\Sigma = -\frac{\nabla t^\star}{\sqrt{-g\left(\nabla t^\star, \nabla t^\star\right)}} = \frac{\sqrt{k_+}}{k_0} \partial_{t^\star} - \frac{2M}{r \sqrt{k_+}}\partial_r \, .
\end{equation}
We also define the (Killing) vectorfields $\Omega_i$ ($i=1,2,3$) to be a basis of generators of the Lie-algebra of $SO\left(3\right)$ corresponding to the spherical symmetry of the Schwarzschild metric. Moreover, we will write $\slashed{\nabla}$ to denote the gradient of the metric induced on the $SO\left(3\right)$-orbits.

Finally, a Penrose diagram of (the black hole exterior of) Schwarzschild-AdS spacetime with two slices of constant $t^\star$ is depicted below.
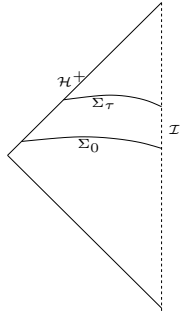
\[
 \input{conformal2.pstex_t}
\]

\subsection{Kerr-AdS}
The Kerr-AdS metric in Boyer Lindquist coordinates reads
\begin{eqnarray}
 g = \frac{\Sigma}{\Delta_-} dr^2 + \frac{\Sigma}{\Delta_\theta} d\theta^2 + \frac{\Delta_\theta\left(r^2+a^2\right)^2 - \Delta_- a^2 \sin^2 \theta}{\Xi^2 \Sigma}\sin^2 \theta d\tilde{\phi}^2 \nonumber \\ -2 \frac{\Delta_\theta \left(r^2+a^2\right) - \Delta_-}{\Xi \Sigma} a \sin^2 \theta d\tilde{\phi} dt - \frac{\Delta_- - \Delta_\theta a^2 \sin^2 \theta}{\Sigma} dt^2
\end{eqnarray}
with the identifications 
\begin{eqnarray}
 \Sigma &=& r^2 + a^2 \cos^2\theta \, , \\
\Delta_\pm &=& \left(r^2+a^2\right)\left(1+\frac{r^2}{l^2}\right) \pm 2Mr \, ,\\
\Delta_\theta &=& 1- \frac{a^2}{l^2} \cos^2 \theta \, ,\\
\Xi &=& 1- \frac{a^2}{l^2} \, .
\end{eqnarray}
Once again, let $k_0 = 1 + \frac{r^2}{l^2}$. A coordinate system which is 
regular on the horizon is obtained by the transformations
\begin{equation}
 t^\star = t + A\left(r\right) \textrm { \ \ \ and \ \ \ } \phi = \tilde{\phi} + B\left(r\right)
\end{equation}
where 
\begin{equation}
 \frac{dA}{dr} = \frac{2Mr}{\Delta_- \left(1+\frac{r^2}{l^2}\right)} \textrm{ \ \ \ and \ \ \ } \frac{dB}{dr} = \frac{a \Xi}{\Delta_-}
\end{equation}
The new metric coefficients become
\begin{eqnarray}
g_{\theta \theta} &=& \frac{\Sigma}{\Delta_\theta} \textrm{ \ \ , \ \ }
g_{t^\star t^\star} = g_{tt}  \textrm{ \ \ , \ \ } \\
g_{\phi \phi} &=& g_{\tilde{\phi} \tilde{\phi}} \textrm{ \ \ , \ \ }
g_{t^\star \phi} = g_{t \tilde{\phi}}  \textrm{ \  , \ \ } \\
g_{rr} &=& \frac{1}{\Sigma \left(1+\frac{r^2}{l^2}\right)^2}\left(\Delta_+ - a^2 \sin^2 \theta \left(k_0 + \frac{\Sigma}{l^2}\right) \right)  \textrm{ \  , \ \ } \\
g_{\phi r} &=& -\frac{a \sin^2 \theta}{\Xi \Sigma k_0} \left[\Xi \Sigma + 2Mr \right]  \textrm{ \  , \ \ } \\
g_{t^\star r} &=& \frac{1}{\Sigma k_0} \left(2Mr - \frac{a^2}{l^2} \sin^2\theta \Sigma \right) \, .
\end{eqnarray}
Note that the angular momentum term in the last expression grows faster in $r$ than the mass term.\footnote{Hence the Kerr-AdS metric is not uniformly close to the Schwarzschild metric in these coordinates! It follows that for the statement that ``Schwarzschild-AdS is close to Kerr-AdS'' one has to use both a regular coordinate patch for $r\leq R$ and a Boyer Lindquist patch for $r>R$ (for some $R$ away from the horizon). For the asymptotically-flat case this is not necessary. \label{coclo}} For $a=0$ the metric reduces to the Schwarzschild-AdS metric in $t^\star, r$ coordinates, cf. (\ref{Schwtr}).
The inverse components are
\begin{eqnarray}
g^{t^\star t^\star} &=& -\frac{\Xi \Delta_+ + a^2 \sin^2 \theta \left(-k_0 \Xi + \frac{2Mr}{l^2}\right)}{k_0^2 \Delta_\theta \Sigma} \, , \\
g^{\theta \theta} &=& \frac{\Delta_\theta}{\Sigma} \textrm{ \ \ , \ \ }
g^{\phi \phi} = \frac{\Xi^2}{\Delta_\theta \Sigma \sin^2 \theta} \textrm{ \ \ , \ \ }
g^{t^\star \phi} = \frac{a \Xi}{k_0 l^2 \Delta_\theta} \, , \\
g^{rr} &=& \frac{\Delta_-}{\Sigma}  \textrm{ \ \ , \ \ \ }
g^{\phi r} = \frac{a \Xi}{\Sigma} \textrm{ \ \ , \ \ \ \ \ \ \ \ \ \ \ \ \ \ }
g^{t^\star r} = \frac{2Mr}{k_0 \Sigma} \, .
\end{eqnarray}
The unit normal of a constant $t^\star$ slice is
\begin{equation}
 n_\Sigma = \sqrt{-g^{t^\star t^\star}} \partial_{t^\star} - \frac{g^{t^\star r}}{\sqrt{-g^{t^\star t^\star}}}\partial_r - \frac{g^{t^\star \phi}}{\sqrt{-g^{t^\star t^\star}}} \partial_\phi \textrm{ \ \ \ , \  $g(n_\Sigma,n_\Sigma)=-1$} \, ,
\end{equation}
with the determinant of the metric induced on constant $t^\star$ slices being
\begin{equation} \label{indmeas}
\sqrt{\det h} = \sqrt{\det g_{t^\star=const}} = \frac{\Sigma}{\Xi} \sin \theta \left(\sqrt{-g^{t^\star t^\star}}\right) \, .
\end{equation}

\section{The dynamics}

\subsection{The class of solutions} \label{WP}
Having defined a proper coordinate system on the black hole backgrounds in the previous section we can turn to the wave equation we would like to study (recall $\Lambda=-\frac{3}{l^2}$):
\begin{equation} \label{wcf}
 \Box_g \psi + \frac{\alpha}{l^2} \psi = 0 \, ,
\end{equation}
for $\alpha \in \mathbb{R}$. The choice $\alpha=2$ in (\ref{wcf}) corresponds to the conformally invariant case, cf.~the discussion in the introduction. Generally we will assume $\alpha < \frac{9}{4}$, the upper Breitenlohner-Freedman bound.

Before we state a theorem on the global dynamics of the above scalar field, let us address the issue of well-posedness of (\ref{wcf}). It turns out that this has not yet been proven explicitly for asymptotically anti de Sitter spacetimes, nor has it for the special cases of Schwarzschild and Kerr-AdS. For pure AdS ($M=a=0$), however, it can be deduced from \cite{Bachelot}. The peculiarity of this problem arises from the fact that due to the timelike nature of $\mathcal{I}$ asymptotically AdS spacetimes are not globally hyperbolic. Hence 
in general ``appropriate'' boundary conditions have to be imposed on $\mathcal{I}$ to make the dynamics wellposed. We will formulate and prove the precise well-posedness statement for asymptotically AdS spacetimes 
in a separate paper. For the purpose of this paper, we content ourselves 
with considering a class of $\psi$ defined as follows:

\begin{definition} \label{wellposed}
Fix $\alpha<\frac{9}{4}$, a Kerr-AdS spacetime $\left(\mathcal{M},g_{M,a,\Lambda}\right)$ and a 
constant $t^\star$ slice $\Sigma_0$ in $\mathcal{D}=\overline{J^+\left(\mathcal{I}\right) \cap J^-\left(\mathcal{I}\right)}$, as well as an integer $k\geq0$. We say that the function $\psi$ is \underline{a solution 
of class $C^k_{dec}$} if in $J^+\left(\Sigma_0\right) \cap J^-\left(\mathcal{I}\right)$ 
\begin{itemize}
 \item $\psi$ is $C^k$
 \item $\psi$ satisfies (\ref{wcf})
 \item for all $\delta<\frac{1}{2}\sqrt{9-4\alpha}$ 
\begin{equation} \label{rdecI}
\lim_{r \rightarrow \infty} |r^{\frac{3}{2}+n+\delta} \partial^n_r \psi| = 0 \textrm{ \ \ \ \ holds for $n=0,1,..., k$} \, .
\end{equation}
\end{itemize}
\end{definition}

In particular, the decay (\ref{rdecI}) ensures that for a solution of class $C^k_{dec}$ 
there is no energy flux (cf.~section \ref{ensec}) through the AdS boundary. 
The decay (\ref{rdecI}) is precisely the one expected from the AdS case \cite{Bachelot} and
also strongly suggested from an asymptotic expansion in $r$ of equation (\ref{wcf}) (performed in \cite{Breitenlohner} for pure AdS). 

The existence of a large class of solutions with the properties of Definition \ref{wellposed}, which arise from appropriate initial data prescribed on $\Sigma_0$, would follow from a general well-posedness statement phrased in terms of weighted Sobolev norms (cf.~also the appendix). We emphasize again that uniqueness is only expected for $\alpha<\frac{5}{4}$.

\subsection{Vectorfield multipliers and commutators}
We will obtain estimates for the field $\psi$ via vectorfield multipliers (and eventually commutators). Since the general technique is well known and reviewed in detail in \cite{Mihalisnotes} we will only give a brief summary. 
The starting point is the energy momentum tensor of the above scalar field
\begin{equation}
T_{\mu \nu} = \partial_\mu \psi \partial_\nu \psi - \frac{1}{2}g_{\mu \nu} \left[\left(\partial_\beta \psi \partial^\beta \psi\right) - \frac{\alpha}{l^2} \psi^2 \right] \, .
\end{equation}
It satisfies
\begin{equation} \label{divo}
\nabla^\mu T_{\mu \nu} = 0
\end{equation}
provided that (\ref{wcf}) holds. Consider a vectorfield $X$ on spacetime. We define its associated currents
\begin{equation}
 J^X_\mu = T_{\mu \nu} X^\nu \,,
\end{equation}
\begin{equation}
\mathbf{K}^X = T_{\mu \nu} \phantom{}^{(X)}\pi^{\mu \nu}
\end{equation}
where  $\phantom{}^{(X)}\pi^{\mu \nu} = \frac{1}{2} \left(\nabla^\mu X^\nu + \nabla^\nu X^\mu\right)$ is the deformation tensor of the vectorfield $X$. One has (using (\ref{divo})) the identity
\begin{equation} \label{vecid}
\nabla^\mu J^X_{\mu} = \nabla^\mu \left( T_{\mu \nu} X^\nu\right) = \mathbf{K}^X \, .
\end{equation}
 Note that $\pi$ vanishes if $X$ is Killing. Integrating (\ref{vecid}) over regions of spacetime relates boundary and volume terms via Stokes' theorem. In particular, for background Killing vectorfields we obtain conservation laws. In writing out the aforementioned integral identities we will sometimes not spell out the measure explicitly (e.g.~equation (\ref{mainid})), it being implicit that the measure is the one induced on the slices (cf.~equation (\ref{indmeas})) or the spacetime measure respectively.

Besides using vectorfields as multipliers we will also use them as commutators. If $X$ is a vectorfield and $\psi$ satisfies (\ref{study}) then $X\left(\psi\right)$ satisfies (cf.~the appendix of \cite{Mihalisnotes})
\begin{equation}
 \Box_g X\left(\psi\right) + \frac{\alpha}{l^2} X\left(\psi\right) = - 2\phantom{}^{(X)}\pi^{\gamma \beta} \nabla_\gamma \nabla_\beta \psi - 2 \left(2\left(\nabla^\gamma \phantom{}^{(X)}\pi_{\gamma \mu}\right) - \nabla_\mu \phantom{}^{(X)} \pi^\gamma_\gamma \right)\nabla^\mu \psi \, . \nonumber
\end{equation}
Note that if $X$ is Killing the right hand side vanishes. In general one may 
apply multipliers to the commuted equation to derive estimates for higher order derivatives. 

\section{Boundedness in the Schwarzschild case}
Here is our boundedness theorem for Schwarzschild-anti de Sitter:
\begin{theorem} \label{maintheo}
Fix a Schwarzschild-anti de Sitter spacetime $\left(\mathcal{M},g_{M > 0,\Lambda}\right)$ and  $\Sigma_0=\Sigma_{\tau_0}$ a slice of constant $t^\star=\tau_0$ in $\mathcal{D}=\overline{J^+\left(\mathcal{I}\right) \cap J^-\left(\mathcal{I}\right)}$.\footnote{Note that such slices satisfy in particular $\mathcal{H}^- \cap \Sigma_0 = \emptyset$.} Let $\alpha<\frac{9}{4}$ and $\psi$ 
be a solution to (\ref{wcf}) of class $C^{n+1}_{dec}$ with $\Omega_i^k \psi \in C^{n+1-k}_{dec}$ for $k=0,...,n$, where $n\geq 0 $ is an integer. If
\begin{equation}
\sum_{k=0}^n \int_{\Sigma_{0}} \left(\frac{1}{r^2} \left(\partial_{t^\star} \Omega^k \psi \right)^2 + r^2 \left(\partial_{r} \Omega^k \psi \right)^2 + |\slashed{\nabla} \Omega^k \psi|^2 \right) r^2 dr d\omega < \infty
\end{equation}
then
\begin{eqnarray}
 \sum_{k=0}^n \int_{\Sigma_{\tau}} \left(\frac{1}{r^2} \left(\partial_{t^\star} \Omega^k \psi \right)^2 + r^2 \left(\partial_{r} \Omega^k \psi \right)^2 + |\slashed{\nabla} \Omega^k \psi|^2 \right) r^2 dr d\omega  \nonumber \\ \leq C \left[\sum_{k=0}^n \int_{\Sigma_{0}} \left(\frac{1}{r^2} \left(\partial_{t^\star} \Omega^k \psi \right)^2 + r^2 \left(\partial_{r} \Omega^k \psi \right)^2 + |\slashed{\nabla} \Omega^k \psi|^2 \right) r^2 dr d\omega\right] \, 
\end{eqnarray}
for a constant $C$ which just depends on $M$, $l$ and $\alpha$. Here $\Sigma_{\tau}$ denotes any constant $t^\star$ slice to the future of $\Sigma_0$ and restricted to $r \geq r_{hoz}$.
\end{theorem}
By Sobolev embedding on $S^2$ we immediately obtain
\begin{corollary} \label{pointw}
The pointwise bound
\begin{equation}
|\psi| \leq C\frac{\sqrt{\left[\sum_{k=0}^2 \int_{\Sigma_{0}} \left(\frac{1}{r^2} \left(\partial_{t^\star} \Omega^k \psi \right)^2 + r^2 \left(\partial_{r} \Omega^k \psi \right)^2 + |\slashed{\nabla} \Omega^k \psi|^2 \right) r^2 dr d\omega\right]}}{r^\frac{3}{2}}
\end{equation}
holds in the exterior $J^-\left(\mathcal{I}^+\right) \cap J^+ \left(\Sigma_0\right)$ for a constant $C$ just depending on the initial data, $M$, $l$ and $\alpha$.
\end{corollary}
\vspace{.4cm}
The remainder of this section is spent proving the above theorem.
In the following we denote by $B$ and $b$ constants which just depend on the fixed parameters $M$, $l$ and $\alpha$. We also define
\begin{equation}
\mathcal{R}\left(\tau_1,\tau_2\right) = \cup_{\tau_1 \leq \tau \leq \tau_2} \Sigma_{\tau}
\end{equation}
to be the region enclosed by the slices $\Sigma_{\tau_1}$ and $\Sigma_{\tau_2}$, a piece of $\mathcal{I}$, and the horizon piece
\begin{equation}
\mathcal{H} \left(\tau_1,\tau_2\right) = \mathcal{H} \cap J^+\left(\Sigma_{\tau_1}\right) \cap J^-\left(\Sigma_{\tau_2}\right) \, .
\end{equation}
Compare the figure above.

\subsection{Positivity of Energy} \label{ensec}
The first step is to obtain a positive energy arising from the Killing vectorfield $T=\partial_{t^\star}$ . For this we apply the vectorfield identity (\ref{vecid}) in the region $\mathcal{R}\left(t^\star_1,t^\star_2\right)$.
For the energy flux through a slice $\Sigma$ we obtain
\begin{eqnarray} \label{Tflux}
E\left(t^\star\right) = \int_{r_{hoz}}^{\infty} \int_{S^2} T\left(\partial_{t^\star}, n_\Sigma\right) \sqrt{\bar{g}_\Sigma} dr d\omega = \nonumber \\  \frac{1}{2} \int_{r_{hoz}}^{\infty} \int_{S^2} \left(\left(\partial_{t^\star} \psi\right)^2 \left[ \frac{k_+}{k_0^2}\right] + \left(\partial_r \psi\right)^2 \left(1-\mu\right) + \left(|\slashed{\nabla} \psi|^2 -\frac{\alpha}{l^2} \psi^2 \right)  \right) r^2 dr d\omega \, .
\end{eqnarray}
The flux through the horizon is 
\begin{equation}
E\left(\mathcal{H}_{[t^\star_1,t^\star_2]}\right) = \int_{\mathcal{H}\left(t^\star_1,t^\star_2\right)} \left(\partial_{t^\star} \psi \right)^2 r^2 dt^\star d\omega \, ,
\end{equation}
hence in particular non-negative. Finally, the flux through $\mathcal{I}$, the AdS boundary, vanishes because of the boundary conditions imposed. Combining these facts we obtain the energy identity
\begin{equation} \label{enid}
E\left(t^\star_2\right) = E\left(t^\star_1\right) + E\left(\mathcal{H}_{[t^\star_1,t^\star_2]}\right) \,
\end{equation}
stating in particular that $E\left(t^\star\right)$ is non-increasing. Next we show that 
the energy flux through the slices $\Sigma$ is positive:
\begin{lemma} \label{posen}
We have
\begin{eqnarray}
E\left(t^\star\right) \geq  \frac{1}{2} \int_{r_{hoz}}^{\infty} \int_{S^2} \Bigg(\left(\partial_{t^\star} \psi\right)^2 \left[ \frac{k_+}{k_0^2}\right] +\left(1-\frac{4}{9}\alpha\right)  \left(\partial_r \psi\right)^2 \left(1-\mu\right) \nonumber \\ + |\slashed{\nabla} \psi|^2\Bigg) r^2 dr d\omega \, .
\end{eqnarray}
\end{lemma}
\begin{proof}
\begin{eqnarray}
\int_{r_{hoz}}^\infty \psi^2 r^2 \ dr = \frac{r^3 - r_{hoz}^3}{3} \psi^2 \Big|^\infty_{r_{hoz}} - \frac{2}{3} \int_{r_{hoz}}^\infty \psi \psi_r \left(r^3 - r_{hoz}^3\right) dr \, .
\end{eqnarray}
The boundary term vanishes because of the decay of $\psi$ at infinity and we obtain the Hardy inequality
\begin{equation}
\int_{r_{hoz}}^\infty \psi^2 r^2 dr \leq \frac{4}{9}  \int_{r_{hoz}}^\infty \left(\partial_r \psi\right)^2 \frac{\left(r^3 - r_{hoz}^3\right)^2}{r^2} dr \leq \frac{4}{9}  \int_{r_{hoz}}^\infty \left(\partial_r \psi\right)^2 r^2 l^2 \left(1-\mu\right) dr 
\end{equation}
where we used (\ref{cubic}) and that $r^3 - r_{hoz}^3 \leq r^3$ in the last step. Hence
\begin{equation} \label{Hardy}
\frac{\alpha}{l^2} \int_{r_{hoz}}^\infty \psi^2 r^2 dr \leq \frac{4}{9} \alpha \int_{r_{hoz}}^\infty \left(\partial_r \psi\right)^2 r^2 \left(1-\mu\right) dr \, .
\end{equation}
Inserting this into (\ref{Tflux}) yields the result.
\end{proof}
The Lemma clearly reveals the relevance of the Breitenlohner-Freedman bound in ensuring a positive energy. We emphasize that one has positivity of energy only in an integrated sense! Lemma \ref{posen} also answers a question posed in \cite{Kunduri} on whether one can construct a positive energy for the 
range $0 < \alpha < \frac{9}{4}$. Finally, we note that a similar Hardy inequality was used 
previously in the appendix of \cite{Bachelot}, where the massive wave equation is studied on pure AdS.

Having established positivity of the energy we only have to deal with the fact that the control over the $\left(\partial_r \psi\right)^2$ term degenerates at the horizon. This is achieved using a so-called ``redshift vectorfield''. See \cite{DafRod2} for its first appearance in the context of asymptotically flat Schwarzschild black holes and the recent \cite{Mihalisnotes} for a version applicable to all non-extremal black holes. An additional difficulty in the present context lies in the fact that the energy momentum tensor does not satisfy the dominant energy condition.

\subsection{The redshift}
Define
\begin{equation}
Y  = \left[\frac{\gamma}{2k_0} + \beta \frac{k_+}{2k_0} \right] \partial_{t^\star} + \left[-\frac{\gamma}{2} + \beta \frac{1-\mu}{2} \right] \partial_r \, .
\end{equation}
Here $\gamma \geq 0$ and $\beta \geq 0$ are both functions of $r$ only, which we will 
define below. We compute the current
\begin{eqnarray}
J^Y_\mu n_\Sigma^\mu = T\left(Y, n_\Sigma\right) = \nonumber \\
\frac{\left(\partial_{t^\star} \psi\right)^2}{2k_0^4} \left[\gamma \left(\sqrt{k_+}\frac{k_0^2}{2}\right) + \beta\left(k_+^\frac{3}{2} \frac{k_0^2}{2}\right)  \right] \nonumber \\
+ \frac{\left(\partial_{t^\star} \psi\right)\left(\partial_{r} \psi\right)}{2k_0} \left[-\sqrt{k_+} \gamma + \beta \sqrt{k_+} k_-\right] \nonumber \\
+ \frac{\left(\partial_{r} \psi\right)^2}{2k_0} \left[\frac{\gamma}{2} \sqrt{k_+} k_0  + \frac{\beta}{2} \frac{k_-^2 k_0}{\sqrt{k_+}}\right] \nonumber \\
+ \frac{|\slashed{\nabla} \psi|^2 - \frac{\alpha}{l^2} \psi^2}{2} \left[\frac{\gamma}{2\sqrt{k_+}} + \frac{\beta}{2} \sqrt{k_+}\right] \, .
\end{eqnarray}
Next consider the current associated with $N=T+e Y$ for some a small constant $e$ (depending on $\alpha$):
\begin{eqnarray}
J^N_\mu n_\Sigma^\mu = T\left(N, n_\Sigma\right) = \nonumber \\
\frac{\left(\partial_{t^\star} \psi\right)^2}{2k_0^4} \left[\sqrt{k_+}k_0^3 + e\gamma \left(\sqrt{k_+}\frac{k_0^2}{2} \right) + e\beta\left(k_+^\frac{3}{2} \frac{k_0^2}{2} \right)  \right] \nonumber \\
+ \frac{\left(\partial_{t^\star} \psi\right)\left(\partial_{r} \psi\right)}{2k_0} \left[-\sqrt{k_+} e\gamma + e \beta \sqrt{k_+} k_-\right] \nonumber \\
+ \frac{\left(\partial_{r} \psi\right)^2}{2k_0} \left[\frac{k_0^2}{\sqrt{k_+}} k_- + e\frac{\gamma}{2} \sqrt{k_+} k_0  + e\frac{\beta}{2} \frac{k_-^2 k_0}{\sqrt{k_+}}\right] \nonumber \\
+ \frac{|\slashed{\nabla} \psi|^2 - \frac{\alpha}{l^2} \psi^2}{2} \left[\frac{k_0}{\sqrt{k_+}} + e\frac{\gamma}{2\sqrt{k_+}} + e\frac{\beta}{2} \sqrt{k_+}\right] \, .
\end{eqnarray}
The bulk term of the vectorfield $N$ reads
\begin{eqnarray}
\mathbf{K}^Y = \frac{1}{e} \mathbf{K}^N = \pi_Y^{\alpha \beta} T_{\alpha \beta} = \nonumber \\
\frac{\left(\partial_{t^\star} \psi\right)^2}{2k_0^2}\left(\left[\gamma \left(\frac{2M}{r^2} + \frac{2r}{l^2} \right) - \gamma_{,r} k_- \right] + k_+^2 \beta_{,r} + 2\frac{k_+}{r} \left(\beta \left(1-\mu\right) - \gamma\right)\right)+\nonumber \\
\frac{\left(\partial_{t^\star} \psi\right)\left(\partial_{r} \psi\right)}{2k_0}\left(-2 \left[\gamma \left(\frac{2M}{r^2} + \frac{2r}{l^2} \right) - \gamma_{,r} k_- \right] + 2\beta_{,r} k_- k_+ - \frac{8M}{r^2}  \left(\beta \left(1-\mu\right) - \gamma\right)\right) \nonumber \\
+{\left(\partial_{r} \psi\right)^2}\left(\gamma \left(\frac{M}{r^2} + \frac{r}{l^2} \right) - \frac{\gamma_{,r} k_-}{2}  + \frac{k_-^2}{2}\beta_{,r} -  \frac{k_-}{r} \left(\beta \left(1-\mu\right) - \gamma\right)\right) \nonumber \\
+|\slashed{\nabla} \psi|^2 \left(-\frac{k_-\beta_{,r}}{2} - \frac{\beta}{2} \left[\frac{2M}{r^2} + \frac{2r}{l^2}\right] + \frac{\gamma_{,r}}{2}\right) \nonumber \\
+ \frac{\alpha}{l^2} \psi^2  \left(\frac{1}{r} \beta k_- - \frac{\gamma}{r} + \frac{k_-\beta_{,r}}{2} + \frac{\beta}{2} \left[\frac{2M}{r^2} + \frac{2r}{l^2}\right] - \frac{\gamma_{,r}}{2} \right) \, .
\end{eqnarray}

Let us define the functions $\gamma$ and $\beta$. Set
$\gamma = \xi\left(r\right) \left(1 + \left(1-\mu\right)\right)$ 
and $\beta = \frac{1}{\delta} \left(1-\mu\right) \xi\left(r\right)$. 
Here $\xi$ is a smooth positive function equal to $1$ in $r\leq r_0$ and identically zero for $r \geq r_1$. The quantity $\delta$ is a small parameter. (Note that $\beta_{,r}$ is positive 
in $r\leq r_0$). Let us regard $\xi$ as being fixed. We choose $r_0$ and $\delta$ such that the 
following conditions hold in $r \leq r_0$
\begin{eqnarray} \label{ineq1}
\left[\gamma \left(\frac{2M}{r^2} + \frac{2r}{l^2} \right) - \gamma_{,r}k_- \right] + {k_+^2} \beta_{,r} + 2\frac{k_+}{r} \left(\beta \left(1-\mu\right) - \gamma\right) \geq b \, ,
\end{eqnarray}
\begin{eqnarray} \label{ineq2}
 \left(\gamma \left(\frac{M}{r^2} + \frac{r}{l^2} \right) - \frac{\gamma_{,r} k_-}{2}  + \frac{k_-^2}{2}\beta_{,r} -  \frac{k_-}{r} \left(\beta \left(1-\mu\right) - \gamma\right)\right) \geq b \, ,
\end{eqnarray}
\begin{eqnarray} \label{ineq3}
\left(-\frac{\beta_{,r}k_-}{2} - \frac{\beta}{2} \left[\frac{2M}{r^2} + \frac{2r}{l^2}\right] + \frac{\gamma_{,r}}{2}\right) \geq b
\end{eqnarray}
and
\begin{eqnarray} \label{ineq4}
\frac{\left(\partial_{t^\star} \psi\right)\left(\partial_{r} \psi\right)}{2k_0}\left(-2 \left[\gamma \left(\frac{2M}{r^2} + \frac{2r}{l^2} \right) - \gamma_{,r}k_- \right] + 2\beta_{,r} k_- k_+ - \frac{8M}{r^2}  \left(\beta \left(1-\mu\right) - \gamma\right)\right) \nonumber \\ \leq \frac{1}{2} \Bigg[\frac{\left(\partial_{t^\star} \psi\right)^2}{2k_0^2}\left(\left[\gamma \left(\frac{2M}{r^2} + \frac{2r}{l^2} \right) - \gamma_{,r}k_- \right] + k_+^2 \beta_{,r} + 2\frac{k_+}{r} \left(\beta \left(1-\mu\right) - \gamma\right)\right) \nonumber \\ + {\left(\partial_{r} \psi\right)^2}\left(\gamma \left(\frac{M}{r^2} + \frac{r}{l^2} \right) - \frac{\gamma_{,r}k_-}{2}  + \frac{k_-^2}{2}\beta_{,r} -  \frac{k_-}{r} \left(\beta \left(1-\mu\right) - \gamma\right)\right)\Bigg] \, .
\end{eqnarray}
It is easily seen all these inequalities can be achieved by inserting the expressions for $\gamma$ and $\beta$, then choosing $\delta$ sufficiently small and finally $r_0$ sufficiently close to $r_{hoz}$ to exploit factors of $k_- = \left(1-\mu\right)$. With $\gamma$ and $\beta$ being determined choose $e$ so small that
\begin{equation} \label{con1}
 e \gamma \leq \frac{1}{2k_0}
\end{equation}
and, in case that $0<\alpha<\frac{9}{4}$, also
\begin{equation} \label{con2}
 e \sup_r \left(\frac{1}{2k_0} \left(\gamma + \beta k_+\right) \right) < \frac{9 \left(1-\frac{4}{9}\alpha\right)}{8\alpha}
\end{equation}
hold for all $r$.

\subsubsection{The $N$ bulk term} 
Let us denote by $\mathbf{K}^N_0$ the expression for $\mathbf{K}^N$ with the zeroth order term removed. 
\begin{lemma} \label{Kbulk}
The quantities $\delta$, $r_0$ and $r_1$ can be chosen such that
\begin{equation}
 \mathbf{K}^N_0 \geq b \left[\left(\partial_{t^\star} \psi\right)^2 + \left(\partial_{r} \psi\right)^2 + |\slashed{\nabla} \psi|^2\right]
\end{equation}
pointwise in $r \leq r_0$. On the other hand,
\begin{equation}
 -\mathbf{K}^N_0 \leq  B \left[\left(\partial_{t^\star} \psi\right)^2 + \left(\partial_{r} \psi\right)^2 + |\slashed{\nabla} \psi|^2\right]
\end{equation}
pointwise in $r_0 \leq r \leq r_1$.
\end{lemma}
\begin{proof}
 The second inequality is immediate since we are away from the horizon. The first is a consequence of the inequalities (\ref{ineq1})-(\ref{ineq4}).

\subsubsection{The $N$ boundary terms} 
Let us investigate what the current $J^N$ actually controls. We can write
\begin{eqnarray}
J^N_\mu n_\Sigma^\mu = T\left(N, n_\Sigma\right) = \nonumber \\
\frac{\left(\partial_{t^\star} \psi\right)^2}{2} \left[\frac{\sqrt{k_+}}{k_0} - e\gamma\frac{\sqrt{k_+}}{2k_0^2} \right] \nonumber \\
 + \frac{\left(\partial_{r} \psi\right)^2}{2} \left[\frac{k_0}{\sqrt{k_+}} k_- + \frac{e \gamma \sqrt{k_+}}{4} \right]  \nonumber \\
+ e\frac{\gamma}{4}\sqrt{k_+} \left[2\left(\frac{\partial_{t^\star} \psi}{k_0} - \frac{\partial_{r} \psi}{2}\right)^2\right] + e\frac{\beta}{4} \sqrt{k_+} \left(\frac{\sqrt{k_+}}{k_0} \partial_{t^\star} \psi + \frac{k_-}{\sqrt{k_+}}\partial_{r} \psi \right)^2 \nonumber \\
+ \frac{|\slashed{\nabla} \psi|^2 - \frac{\alpha}{l^2} \psi^2}{2} \left[\frac{k_0}{\sqrt{k_+}} + e\frac{\gamma}{2\sqrt{k_+}} + e\frac{\beta}{2} \sqrt{k_+}\right]
\end{eqnarray}
with the term in the penultimate line being manifestly non-negative. From (\ref{con1}) one easily obtains a lower bound for the square bracket multiplying the $\left(\partial_{t^\star} \psi\right)^2$-term. Inequality (\ref{con1}) in turn allows us to control the bad (for $0<\alpha<\frac{9}{4}$ !) zeroth order term by borrowing from the $r$-derivative term using the Hardy inequality (\ref{Hardy}). Namely, if $0<\alpha<\frac{9}{4}$, then:
\begin{eqnarray}
\int_{r_{hoz}}^\infty \frac{\alpha}{l^2} \psi^2 \left[1 + \frac{e}{2k_0} \left(\gamma + \beta k_+\right)\right] r^2 dr \nonumber \\ \leq \left(1 + \frac{9 \left(1-\frac{4}{9}\alpha\right)}{8\alpha}\right) \int_{r_{hoz}}^\infty \frac{\alpha}{l^2} \psi^2 r^2 dr \nonumber \\  \leq \frac{1}{2}\left(1+\frac{4}{9}\alpha\right) \int_{r_{hoz}}^\infty \left(\partial_r \psi\right)^2 \left(1-\mu\right) r^2 dr \, .
\end{eqnarray}
Hence finally
\begin{eqnarray}
\int_{r_{hoz}}^\infty \int_{S^2} J^N_\mu n_\Sigma^\mu \frac{\sqrt{k_+}}{k_0^2} r^2 dr d\omega \geq \\ 
\int_{r_{hoz}}^\infty \int_{S^2} \frac{\left(\partial_{t^\star} \psi\right)^2}{2} \left[\frac{\sqrt{k_+}}{2k_0}  \right] \nonumber \\
+ \frac{\left(\partial_{r} \psi\right)^2}{2} \left[\frac{1}{2} \left(1-\frac{4}{9}\alpha\right) \left(1-\mu\right) + e\frac{\gamma \sqrt{k_+}}{4} \right] r^2 dr d\omega \nonumber \\
+ \int_{r_{hoz}}^\infty \int_{S^2} \frac{|\slashed{\nabla} \psi|^2}{2}  r^2 dr d\omega 
\end{eqnarray}
and we have established the following
\begin{lemma}
\begin{eqnarray}
\int_{\Sigma} J^N_\mu n_\Sigma^\mu = \int_{r_{hoz}}^\infty \int_{S^2} J^N_\mu n_\Sigma^\mu \frac{\sqrt{k_+}}{k_0} r^2 dr d\omega \geq \nonumber \\ b \int_{r_{hoz}}^\infty \int_{S^2} \left[\frac{\left(\partial_{t^\star} \psi\right)^2}{r^2} + \left(\partial_{r} \psi\right)^2 r^2 + |\slashed{\nabla} \psi|^2 \right] r^2 dr d\omega \, .
\end{eqnarray} 
\end{lemma}
Note that the degeneration of the $\left(\partial_{r} \psi\right)^2$-term which occurred in the $T$-energy has disappeared.

\subsection{The boundedness}

\begin{proposition} \label{rsest}
There exist constants $B$ and $b$ such that for any $\tau_2 \geq \tau_1 \geq 0$
\begin{eqnarray} \label{cru}
- \int_{\mathcal{H}\left(\tau_1,\tau_2\right)} J^N_\mu n^\mu_{\mathcal{H}^+} + b \int_{\tau_1}^{\tau_2} d\tau \int_{\Sigma_\tau} J^N_\mu n^\mu_{\Sigma} \nonumber \\ \leq \int_{\mathcal{R}\left(\tau_1, \tau_2\right)} \mathbf{K}^N + B  \int_{\tau_1}^{\tau_2} d\tau  \int_{\Sigma_\tau}  J^T_\mu n^\mu_{\Sigma} \, .
\end{eqnarray}
\end{proposition}
\begin{proof}
We first note that all zeroth order terms can be absorbed by the last term in (\ref{cru}) using inequality (\ref{Hardy}). Hence it suffices to prove the inequality for the first order terms.
We have
\begin{equation} \label{hozterm}
 \int_{\mathcal{H}\left(\tau_1,\tau_2\right)} J^N_\mu n^\mu_{\mathcal{H}^+} = \int_{\tau_1}^{\tau_2} \int_{S^2} \left[e \gamma \frac{M}{2r k_0} \left(|\slashed{\nabla} \psi|^2 - \frac{\alpha}{l^2} \psi^2\right) + \left(\partial_{t^\star} \psi\right)^2\right] r^2 dt^\star d\omega \, .
 \end{equation}
For $\alpha \leq 0$ the term has a good sign while for $0<\alpha<\frac{9}{4}$ 
 \begin{eqnarray} \label{hoz2}
- \int_{\mathcal{H}\left(\tau_1,\tau_2\right)} J^N_\mu n^\mu_{\mathcal{H}^+} \leq \int_{\tau_1}^{\tau_2} dt^\star \int_{S^2} d\omega \int_{r_{hoz}}^{r_1} dr \partial_r \left[-e \gamma \frac{Mr}{2 k_0} \left(\frac{\alpha}{l^2} \psi^2\right)\right] \nonumber \\
\leq e \int_{\mathcal{R}\left(\tau_1, \tau_2\right) \cap \{r \leq r_1\}}  \left( \eta\left( \partial_{r} \psi\right)^2 + \frac{B}{\eta}\frac{\alpha}{l^2} \psi^2\right) dt^\star dr d\omega
 \end{eqnarray}
for any $\eta>0$ using that $\gamma$ is supported for $r < r_1$ only. So we only need a little bit of the $\partial_r$ term which we can borrow from the good bulk term $\mathbf{K}^N$ in $r\leq r_0$ (by Lemma \ref{Kbulk}) and from the $T$-energy term in the remaining region. We have
\begin{eqnarray}
\int_{\mathcal{R}\left(\tau_1, \tau_2\right)\cap \{r \leq r_1\}} \left( -\mathbf{K}^N +  \left( e \eta\left( \partial_{r} \psi\right)^2 + e\frac{B}{\eta}\frac{\alpha}{l^2} \psi^2\right) \right) + b \int_{\tau_1}^{\tau_2} d\tau \int_{\Sigma_\tau} J^N_\mu n^\mu_{\Sigma}  \nonumber \\ \leq \tilde{B}  \int_{\tau_1}^{\tau_2} d\tau  \int_{\Sigma_\tau}  J^T_\mu n^\mu_{\Sigma} 
\end{eqnarray}
for a small constant $b$ and large constants $B$ and $\tilde{B}$ as a consequence of 
Lemma \ref{Kbulk}. Combining this inequality with (\ref{hoz2}) yields (\ref{cru}).
\end{proof}

{\bf Remark:} We have discarded good terms (the $t^\star$ and angular derivative) on the 
horizon in the proof of the proposition as they are not needed for the following argument. Later, when we start commuting the equation with the redshift vectorfield we will need to keep those positive terms in order to estimate certain errorterms (cf.~section \ref{ho+pw}).

Using the previous Proposition we can prove boundedness as follows. The $N$ identity reads
\begin{eqnarray} \label{mainid}
 \int_{\Sigma_\tau} J^N_\mu n^\mu_{\Sigma} +  \int_{\mathcal{H}^+} J^N_\mu n^\mu_{\mathcal{H}^+} + \int_{\mathcal{R}\left(0,\tau\right)} \mathbf{K}^N = \int_{\Sigma_\tau} J^N_\mu n^\mu_{\Sigma_0} \, .
\end{eqnarray}
An application of Proposition \ref{rsest} yields the inequality
\begin{eqnarray}
 \int_{\Sigma_{\tau_2}} J^N_\mu n^\mu_{\Sigma} + b \int_{\tau_1}^{\tau_2} d\tau \int_{\Sigma_\tau} J^N_\mu n^\mu_{\Sigma} \leq B  \int_{\tau_1}^{\tau_2} d\tau  \int_{\Sigma_\tau}  J^T_\mu n^\mu_{\Sigma} + \int_{\Sigma_{\tau_1}} J^N_\mu n^\mu_{\Sigma_{\tau_1}} \, .
\end{eqnarray}
Using (\ref{enid}) (in particular the fact that the $T$-energy is non-increasing from initial data) and setting $f \left(\tau\right) = \int_{\Sigma_\tau} J^N_\mu n^\mu_{\Sigma} $ as well as $D= \int_{\Sigma_0}  J^T_\mu n^\mu_{\Sigma} $ we arrive at
\begin{eqnarray} \label{basic}
f\left(\tau_2\right) + b \int_{\tau_1}^{\tau_2} f\left(\tau\right)  d\tau \leq B D \left(\tau_2-\tau_1\right) + f\left(\tau_1\right)
\end{eqnarray}
for any $\tau_2 \geq \tau_1 \geq 0$ from which boundedness of $f\left(\tau\right)$ follows from a pigeonhole argument, cf. \cite{Mihalisnotes}. 

Due to the spherical symmetry of the background we can commute equation (\ref{wcf}) with angular momentum operators and obtain boundedness of
\begin{equation}
\int_{\Sigma} J^N_\mu \left(\Omega^k \psi\right) n^\mu_\Sigma 
\end{equation}
for any integer $k$ assuming such a bound on the data. Theorem \ref{maintheo} follows.
\end{proof}

\subsection{Pointwise bounds}
For completeness we give the proof of Corollary \ref{pointw}. We have
\begin{eqnarray}
 |\psi\left(t^\star, r, \omega\right)| \leq \int_r^\infty |\partial_r \psi\left(t^\star, r, \omega\right) | dr \leq \frac{1}{\sqrt{3}r^\frac{3}{2}} \sqrt{\int_r^\infty |\partial_r \psi\left(t^\star, r, \omega\right) |^2 r^4 dr} \, ,
\end{eqnarray}
and by Sobolev embedding on $S^2$
\begin{eqnarray}
\int_r^\infty |\partial_r \psi\left(t^\star, r, \omega\right) |^2 r^4 dr \leq 
\tilde{C} \int_r^\infty \int_{S^2} \sum_{k=0}^2 \left(|\partial_r \Omega^k\left(\psi\right)\left(t^\star, r, \omega\right) |^2\right) r^4 dr d\omega_{S^2}  \nonumber \\ \leq \tilde{C} \sum_{k=0}^2 \int_{\Sigma} J^N_\mu \left(\Omega^k \psi\right) n^\mu_\Sigma  \, ,
\end{eqnarray}
from which the statement of the corollary follows. 

\subsection{Higher order quantities}
Clearly, via commuting with $T$ and angular momentum operators, we obtain control over certain higher order energies as well. Using just this collection of vectorfields however, we will still not be able to estimate the derivative transverse to the horizon. This problem was resolved for the asymptotically-flat Kerr case in \cite{DafRodKerr, Mihalisnotes}, roughly speaking by commuting with the redshift vectorfield as well. We will adapt their argument to the present asymptotically hyperbolic case in section \ref{ho+pw}, when we deal with the Kerr-AdS metric. There the argument is unavoidable even to obtain a pointwise bound on $\psi$ since one can no longer (trivially) commute with angular momentum operators!

\section{The Kerr-AdS case} \label{KerrAdS}
To generalize the argument to include the case of Kerr-AdS we have to circumvent several difficulties. First of all, the timelike Killing field $T=\partial_{t^\star}$ is no longer timelike everywhere on the black hole exterior due to the presence of an ergoregion close to the horizon. Hence the $T$-identity alone will not produce positive boundary terms. The resolution is to consider the Killing field
\begin{equation}
 K = T + \lambda \Phi
\end{equation}
for an appropriate constant $\lambda$ and $\Phi=\partial_\phi$ being the Killing field corresponding to the axisymmetry of the Kerr-AdS metric. $K$ is everywhere timelike on the black hole exterior (it is null on the horizon, coinciding with the generators) as long as $|a|l < r_{hoz}^2$. This is a consequence of the asymptotically AdS nature of the background\footnote{more precisely, the fact that $g_{t^\star t^\star} \sim g_{t^\star \phi} \sim g_{\phi \phi} \sim r^2$}: Note that the analogous vectorfield in asymptotically flat space would turn spacelike near infinity! These properties of $K$ were first exploited by Hawking and Reall \cite{HawkingReall} to obtain a pointwise positive energy for fields whose energy momentum tensor satisfies the dominant energy condition, see the discussion in the introduction. Not surprisingly, we will show below that for sufficiently small $a$ the energy identity associated to $K$ produces boundary terms with manifestly non-negative first order terms. The zeroth order term however still has the wrong sign for $0<\alpha<\frac{9}{4}$ (as in Schwarzschild-AdS), a consequence of the dominant energy condition being violated in this range. We resolve this issue by generalizing the Hardy inequality (\ref{Hardy}) to the Kerr-AdS case allowing us to again control the zeroth order term from the derivative term in the $K$ energy. Because we have to ``borrow'' from the $r$-derivative term for the Hardy inequality, we have to impose stronger restrictions on the smallness of $a$ than just $|a|l < r_{hoz}^2$.

Since the argument close to the horizon involving the vectorfield $Y$ is stable in itself (it applies to any black hole event horizon with positive surface gravity by Theorem 7.1 of \cite{Mihalisnotes}) our previous argument using the vectorfield $N=K+eY$ immediately yields boundedness of the $L^2$-norm of all derivatives. In particular, no further restrictions on the size of $a$ arise.

The only remaining difficulty is that we can no-longer commute with angular momentum operators to obtain pointwise bounds because the spherical symmetry of Schwarzschild has been broken. The way around this is to commute (\ref{wcf}) with $K$ and $Y$ and to derive elliptic estimates on the asymptotically hyperbolic slices $\Sigma$. This is carried out in section \ref{ho+pw}.

\subsection{The vectorfield $K$: Positive Energy}

We first compute the currents associated with the Killing vectors $T=\partial_{t^\star}$ and $\Phi = \partial_{\phi}$. 
\begin{eqnarray}
 T\left(\partial_{t^\star},n_{\Sigma}\right) = \frac{1}{2} \sqrt{-g^{t^\star t^\star}} \left(\partial_{t^\star} \psi \right)^2 + \frac{1}{2} \frac{\Delta_-}{\Sigma \sqrt{-g^{t^\star t^\star}}}\left(\partial_r \psi\right)^2 - \frac{\alpha}{l^2} \frac{1}{2\sqrt{-g^{t^\star t^\star}}} \psi^2 \nonumber \\ + \frac{1}{2\sqrt{-g^{t^\star t^\star}}} \left(g^{\phi \phi} \left(\partial_\phi \psi\right)^2 + g^{\theta \theta}\left(\partial_\theta \psi\right)^2 + 2g^{r \phi} \left(\partial_r \psi\right) \left(\partial_\phi \psi\right) \right)
\end{eqnarray}
and
\begin{eqnarray}
 T\left(\partial_{\phi},n_{\Sigma}\right) = \sqrt{-g^{t^\star t^\star}} \left[\left(\partial_\phi \psi\right)\left(\partial_{t^\star} \phi\right) + \frac{g^{t^\star r}}{g^{t^\star t^\star}}\left(\partial_\phi \psi\right)\left(\partial_{r} \phi\right) + \frac{g^{t^\star \phi}}{g^{t^\star t^\star}}\left(\partial_\phi \psi\right)^2\right] \, .
\end{eqnarray}
We observe that we cannot control the $\left(\partial_r \psi\right)\left(\partial_\phi \psi\right)$ term in the $T$-energy, because the $\left(\partial_r \psi\right)^2$ term from which we have to borrow degenerates on the horizon. However, consider the vectorfield
\begin{equation}
K = T + \lambda \Phi \textrm{ \ \ \ \ \ with $\lambda = \frac{a \Xi}{r_{hoz}^2+a^2}$ } 
\end{equation}
and its current on constant $t^\star$ slices
\begin{eqnarray} \label{Kcurr}
2 \sqrt{-g^{t^\star t^\star}} \cdot T\left(K,n_{\Sigma}\right) = -\frac{\alpha}{l^2}\psi^2 + g^{\theta \theta}\left(\partial_\theta \psi\right)^2
 \nonumber \\ + \Bigg[\left(-g^{t^\star t^\star}\right)\left(\partial_{t^\star} \psi \right)^2 -g^{t^\star t^\star} \frac{2a \Xi}{r_{hoz}^2+a^2}\left(\partial_\phi \psi\right)\left(\partial_{t^\star} \psi\right) \nonumber \\ + \left(g^{\phi \phi} - g^{t^\star \phi} \frac{2a \Xi}{r_{hoz}^2+a^2}\right)\left(\partial_\phi \psi\right)^2 \Bigg]\nonumber \\
+ \frac{\Delta_-}{\Sigma}\left(\partial_r \psi\right)^2  +2 \left(\partial_r \psi\right) \left(\partial_\phi \psi\right) \left[g^{r \phi} - g^{t^\star r} \frac{a \Xi}{r_{hoz}^2+a^2}\right] \, .
\end{eqnarray}
We choose $a$ so small that
\begin{equation} \label{co1}
\frac{1}{4} g^{\phi \phi} - g^{t^\star \phi} \frac{2a \Xi}{r_{hoz}^2+a^2} = \frac{\Xi^2}{4 \Delta_\theta \Sigma \sin^2 \theta} -\frac{2a^2 \Xi^2}{\left(r_{hoz}^2+a^2\right) k_0 l^2 \Delta_\theta} \geq 0 \, .
\end{equation}
This is possible because both terms decay like $\frac{1}{r^2}$ in $r$. In the asymptotically flat case the $g^{t^\star \phi}$-term will eventually dominate because $k_0=1$ in the asymptotically flat case ($k_0 = 1+\frac{r^2}{l^2}$ in AdS!). We also choose $a$ small enough so that
\begin{eqnarray} \label{co2}
 \frac{1}{2}\left(-g^{t^\star t^\star}\right)\left(\partial_{t^\star} \psi \right)^2 -g^{t^\star t^\star} \frac{2a \Xi}{r_{hoz}^2+a^2}\left(\partial_\phi \psi\right)\left(\partial_{t^\star} \psi\right) + \frac{1}{8} g^{\phi \phi} \left(\partial_\phi \psi\right)^2 \geq 0 \, .
\end{eqnarray}
This is easily achieved since again all terms decay like $\frac{1}{r^2}$ at infinity and the cross-term has a factor of $a$. Of the first order terms it remains to control the $r\phi$ cross-term. For this we note (cf.~Lemma \ref{dmc})
\begin{eqnarray} \label{borpr}
 g^{r \phi} - g^{t^\star r} \frac{a \Xi}{r_{hoz}^2+a^2} &=& \frac{a \Xi}{k_0\left(r_{hoz}^2+a^2\right)\Sigma} \left(\left(r_{hoz}^2+a^2\right)k_0-2Mr\right) \nonumber \\
&=& \frac{a \Xi}{k_0\left(r_{hoz}^2+a^2\right)\Sigma} \left(\Delta_- -k_0 \left(r_{hoz}+r\right)\left(r-r_{hoz}\right) \right) \nonumber \\ 
&=& \frac{a \Xi}{k_0\Sigma} \frac{\left(r-r_{hoz}\right)}{l^2} \left(r-\frac{l^2}{r_{hoz}}\right)  \,.
\end{eqnarray}
The reason for the above choice of $K$ is now obvious: The $\left(\partial_r \psi\right)\left(\partial_\phi \psi\right)$-term has acquired a weight which degenerates on the horizon. Hence we can borrow from the $\left(\partial_r \psi\right)^2$ term whose weight also degenerates. Moreover, the decay of (\ref{borpr}) in $r$ is strong, which will be exploited soon.

Before we continue we obtain rather precise control over how the quantity $\Delta_-$ deteriorates 
on the horizon:
\begin{lemma} \label{dmc}
 We have
\begin{equation}
 \Delta_- = \frac{1}{l^2} \left(r-r_{hoz}\right)\left(r^3 + r^2 r_{hoz} + r \left(r_{hoz}^2+l^2+a^2\right) - \frac{a^2 l^2}{r_{hoz}}\right) \nonumber
\end{equation}
\end{lemma}
and
\begin{equation}
\Delta_- - \frac{\left(r^3-r_{hoz}^3 + 3a^2\left(r-r_{hoz}\right)\right)^2}{\left(r^2+a^2\right)l^2} = f
\end{equation}
where
\begin{eqnarray} \label{fdefn}
f = \frac{1}{\left(r^2+a^2\right)l^2}\left(r-r_{hoz}\right) \Bigg[r^3 \left(l^2-4a^2\right) + r^2 \left(r_{hoz}^3-\frac{a^2 l^2}{r_{hoz}} + a^2 r_{hoz}\right) \nonumber \\ + r\left(r_{hoz}^4-8 a^4 + a^2 \left[r_{hoz}^2+l^2\right]\right) + r_{hoz}\left[r_{hoz}^2+3a^2\right]^2 - \frac{a^4 l^2}{r_{hoz}}\Bigg] \, .
\end{eqnarray}
\begin{proof}
Direct computation.
\end{proof}
\begin{corollary}
For sufficiently small $a$ we have $f \geq 0$.
\end{corollary}
\begin{proof}
 All coefficients of the polynomial in the square bracket of (\ref{fdefn}) are positive for small enough $a$ (the condition $|a|<\frac{1}{2}l$ in addition to $|a| l <r_{hoz}^2$ is seen to be sufficient).
\end{proof}
\begin{corollary} \label{dmc2}
For sufficiently small $a$ we have
\begin{equation}
\frac{16\Sigma}{g^{\phi \phi}} \left(\frac{a \Xi}{k_0\Sigma} \frac{\left(r-r_{hoz}\right)}{l^2} \left(r-\frac{l^2}{r_{hoz}}\right)\right)^2 < f \, .
\end{equation}
\end{corollary}
\begin{proof}
Direct computation.
\end{proof}
Of course we recover our previous result (\ref{cubic}) for $a=0$. Note also 
that $f$ grows slower (like $\sim r^2$) than $\Delta_-$ (which grows like $\sim r^4$). Going back to (\ref{borpr}) we have the estimate
\begin{eqnarray} \label{co3}
 2 \left(\partial_r \phi\right) \left(\partial_\phi \psi\right) \left[g^{r \phi} - g^{t^\star r} \frac{a \Xi}{r_{hoz}^2+a^2}\right] \nonumber \\ \leq \left(\frac{8}{g^{\phi \phi}} \left(\frac{a \Xi}{k_0\Sigma} \frac{\left(r-r_{hoz}\right)}{l^2} \left(r-\frac{l^2}{r_{hoz}}\right)\right)^2 \right) \left(\partial_r \psi\right)^2+ \frac{1}{8} g^{\phi \phi} \left(\partial_\phi \psi\right)^2 \nonumber \\ \leq \frac{f}{2} \frac{\left(\partial_r \phi\right)^2}{\Sigma} + \frac{1}{8} g^{\phi \phi} \left(\partial_\phi \psi\right)^2
\end{eqnarray}
for sufficiently small $a$ using Corollary \ref{dmc2}. It is here where the good decay of (\ref{borpr}) is exploited, allowing us to borrow from $f$ (and not the full $\Delta_-$) only.

Now that positivity has been obtained for all first order terms we can turn to the zeroth order term and generalize the Hardy inequality (\ref{Hardy}). We note
\begin{eqnarray}
 \frac{\alpha}{l^2} \int_{S^2}d\theta d\phi  \int_{r_{hoz}}^\infty dr \sin \theta \frac{\Sigma}{\Xi} \left(\psi\left(t^\star, r, \theta,\phi\right)\right)^2 \nonumber \\ \leq \frac{\alpha}{\Xi l^2}   \int_{S^2}d\theta d\phi \sin \theta \int_{r_{hoz}}^\infty dr \left(r^2+a^2\right) \psi^2
\end{eqnarray}
and estimate as previously,
\begin{eqnarray}
 \frac{\alpha}{l^2} \int_{r_{hoz}}^\infty dr \left(r^2+a^2\right) \psi^2 &\leq& \frac{4}{9} \alpha \int_{r_{hoz}}^\infty dr  \frac{\left(r^3-r_{hoz}^3 + 3a^2 \left(r-r_{hoz}\right)\right)^2}{\left(r^2+a^2\right) l^2}\left(\partial_r \psi\right)^2 \nonumber \\ &=& \frac{4}{9} \alpha \int_{r_{hoz}}^\infty dr  \left(\Delta_- -f\right) \left(\partial_r \psi\right)^2 
\end{eqnarray}
using Lemma \ref{dmc}. We have obtained the Hardy inequality
\begin{eqnarray} \label{KerrHardy}
 \frac{\alpha}{l^2}  \int_{S^2} d\theta d\phi \int_{r_{hoz}}^\infty dr \sin \theta \frac{\Sigma}{\Xi} \left(\psi\left(t^\star, r, \theta,\phi\right)\right)^2 \nonumber \\ \leq \frac{4}{9} \alpha \int_{S^2}d\theta d\phi \int_{r_{hoz}}^\infty dr \frac{\sin \theta}{\Xi}  \left(\Delta_- - f\right) \left(\partial_r \psi\right)^2 \, .
\end{eqnarray}
Inserting the estimates (\ref{co1}), (\ref{co2}), (\ref{co3}) and (\ref{KerrHardy}) into (\ref{Kcurr}) we finally have
\begin{proposition}
For sufficiently small $a$ the estimate
\begin{eqnarray} 
\int_{\Sigma} J^K_{\mu}\left(\psi\right) n^\mu_{\Sigma} =  \int_{S^2}d\theta d\phi \int_{r_{hoz}}^\infty dr \sin \theta \frac{\Sigma}{\Xi} \sqrt{-g^{t^\star t^\star}} J^K_{\mu} n^\mu_{\Sigma} \nonumber \\ 
\geq b\int_{S^2} d\theta d\phi\int_{r_{hoz}}^\infty dr \Sigma \ \sin \theta  \Bigg(\frac{1}{r^2} \left(\partial_{t^\star} \psi \right)^2 + \frac{\Delta_-}{r^2} \left(\partial_{r} \psi\right)^2 \nonumber \\ + \frac{\Xi}{\Delta_\theta \Sigma \sin^2 \theta} \left(\partial_\phi \psi\right)^2 + \frac{\Delta_\theta}{\Sigma \Xi} \left(\partial_{\theta} \psi\right)^2\Bigg)
\end{eqnarray}
holds for a constant $b$ which just depends on $M$, $l$ and $\alpha<\frac{9}{4}$.
\end{proposition}
How large is $a$ allowed to be? Note that in the case $\alpha \geq 0$ our estimates reduce to the simple geometric condition that $K$ is everywhere timelike outside the black hole. This in turn translates into a single inequality for $a$, thereby recovering $|a|l<r_{hoz}^2$, the result of \cite{HawkingReall}. If on the other hand $0<\alpha<\frac{9}{4}$, the additional restriction of Corollary \ref{dmc2} has to be imposed. The reason is that the $\Delta_-\left(\partial_r \phi\right)^2$-term in the $K$-identity has to control both the mixed term (\ref{co3}) and the zeroth order term (\ref{KerrHardy}). We used a coefficient $f$ (independent of $\alpha$) to control the former and $\left(\Delta_- -f\right)$ to control the latter establishing that a uniform $a$ can be chosen for $\alpha<\frac{9}{4}$.

Finally we compute the flux term on the horizon, where the normal is proportional to $K$:
\begin{equation}
\int_{\mathcal{H}} T\left(K,K\right) = \int_{\mathcal{H}} \left(\partial_{t^\star} \psi + \frac{a \Xi}{r_{hoz}^2 + a^2} \partial_{\phi} \psi\right)^2 > 0 \, .
\end{equation}

\subsection{The vectorfield N}
Now that we again have a positive, non-increasing $K$-energy we can invoke the identical argument as in the Schwarzschild case close to the horizon, involving the vectorfields $Y$ and $N=K+eY$. This will eventually produce a bound
\begin{eqnarray}
\int_{S^2} d\theta d\phi \int_{r_{hoz}}^\infty dr \Sigma \ \sin \theta  \Bigg(\frac{1}{r^2} \left(\partial_{t^\star} \psi \right)^2 + r^2 \left(\partial_{r} \psi\right)^2 + \frac{\Xi}{\Delta_\theta \Sigma \sin^2 \theta} \left(\partial_\phi \psi\right)^2 \nonumber \\ + \frac{\Delta_\theta}{\Sigma \Xi} \left(\partial_{\theta} \psi\right)^2\Bigg) \leq B \int_{\Sigma} J^N_{\mu}\left(\psi\right) n^\mu_{\Sigma} \leq B \int_{\Sigma_0} J^N_{\mu}\left(\psi\right) n^\mu_{\Sigma_0} \, ,
\end{eqnarray}
where the ``bad'' $\Delta_-$-weight from the $r$-derivative term has now disappeared. Note that no further smallness restrictions on $a$ arise as long as the horizon has positive surface gravity.
\subsection{Higher order energies and pointwise bounds} \label{ho+pw}
We finally address the issue of pointwise bounds.  Unfortunately, we can no longer exploit the commutation with the angular derivatives because of the broken spherical symmetry. However, we can certainly commute with $T$ (or $K$ respectively) to obtain integral bounds for certain higher order energies. Given a solution of class $C^m_{dec}$ the inequality
\begin{eqnarray} \label{Tcomcon}
\int_{\Sigma} J^N_{\mu} \left(K^k\psi\right) n^\mu_{\Sigma} \leq B \int_{\Sigma_0} J^N_{\mu} \left(K^k\psi\right) n^\mu_{\Sigma} 
\end{eqnarray}
for any non-negative integer $k<m$ is an automatic consequence of the fact that $K$ commutes with the wave operator. 
%
%
%
%
%
%
We write the wave equation as
\begin{eqnarray} \label{wofw}
\frac{1}{\sqrt{g}}\partial_i \left(g^{ij} \sqrt{g} \partial_j \psi \right) \nonumber \\ = -g^{t^\star t^\star}\left(\partial_{t^\star} \partial_{t^\star} \psi \right) - 2g^{t^\star i} \partial_{t^\star} \partial_i \psi - \frac{1}{\sqrt{g}} \partial_r \left(g^{t^\star r} \sqrt{g}\right) \left(\partial_{t^\star} \psi\right) -\frac{\alpha}{l^2}\psi \, ,
\end{eqnarray}
where $i=1,2,3$ (or $r,\theta,\phi$ respectively). Using both the decay assumed for $\psi$ at infinity and that we control the right hand side by (\ref{Tcomcon}), we obtain for any $r_0>r_{hoz}$ the elliptic estimate
\begin{eqnarray} \label{elliptic1}
\int_{\Sigma \cap \{r \geq r_0 > r_{hoz}\}} \frac{1}{\sqrt{-g^{t^\star t^\star}}} h^{ij} h^{kl} \left(\nabla_i \nabla_k \psi \right) \left(\nabla_j \nabla_l \psi \right) \frac{\Sigma}{\Xi}\sin \theta \sqrt{-g^{t^\star t^\star}} dr d\theta d\phi \nonumber \\ 
\leq C \left(r_0\right)\int_{ \Sigma \cap \{r \geq r_{hoz}\}} \left(J^N_{\mu}\left(\psi\right) n^\mu_{\Sigma} + J^N_{\mu}\left( K \psi\right) n^\mu_{\Sigma}\right) \frac{\Sigma}{\Xi}\sin \theta \sqrt{-g^{t^\star t^\star}} dr d\theta d\phi
\end{eqnarray}
away from the horizon, where $h^{ij}$ is the inverse of the induced metric on $\Sigma$. Note that the argument breaks down at the horizon because 
of the degenerating weight of $g^{rr}$ (whereas $h^{rr}$ is well-behaved there).\footnote{Actually, the estimate (\ref{elliptic1}) remains true if one inserts an additional weight of $r^{n}$ for any $n<\max\left(2,\sqrt{9-4\alpha}\right)$ into the integrand on the left hand side. This improvement is outlined in the appendix where stronger weighted Sobolev norms are derived.}

To obtain good estimates close to the horizon we again adapt the ideas of \cite{DafRodKerr, Mihalisnotes}. Their resolution is to commute the equation with a version of the redshift vector field
\begin{equation}
 \hat{Y} = \frac{1}{2k_0} \partial_{t^\star} - \frac{1}{2}\partial_r
\end{equation}
which yields the equation
\begin{equation} \label{commuteY}
 \Box \left(\hat{Y} \psi \right) = +\frac{2\Delta_-^\prime}{\Sigma} \hat{Y}\hat{Y} \left(\psi\right) - \frac{4r}{\Sigma} \hat{Y}T\left(\psi\right) + \lambda_1 \left(\partial_r \psi\right) + \lambda_2 \left(\partial_{t^\star} \psi\right) + \frac{\alpha}{l^2} \hat{Y}\left(\psi\right) \, ,
\end{equation}
where
\begin{equation}
 \lambda_1 = \frac{1}{2\Sigma} \Delta_-^{\prime \prime} \sim \frac{1}{r^0} \, ,
\end{equation}
\begin{equation}
 \lambda_2 = -\frac{a^2+5r^2}{l^2 k_0 \Sigma} \sim \frac{1}{r^2} \, 
\end{equation}
and a prime denotes taking a derivative with respect to $r$. What will be crucial is that the coefficient of the first term on the right hand side of (\ref{commuteY}) has a (good) sign close to the horizon. There is a geometric reason for this which was observed in \cite{Mihalisnotes}: Computing the surface gravity $\kappa$ defined by (cf.~\cite{Wald})
\begin{equation}
 \partial_b \left(K_a K^a\right)\Big|_{r=r_{hoz}} = -2 \kappa K_b \Big|_{r=r_{hoz}}
\end{equation}
we obtain
\begin{equation}
 \kappa = \frac{\Delta_-^\prime \left(r_{hoz}\right)}{2\left(r_{hoz}^2+a^2\right)}
\end{equation}
for the Kerr-AdS black hole under consideration. This means that on the horizon the coefficient of the $\hat{Y}\hat{Y} \left(\psi\right)$-term on the right hand side of (\ref{commuteY}) is proportional to the surface gravity of the black hole, which is positive for a non-degenerate horizon! Remarkably, this observation is not bound to the special properties of the Kerr metric but a general fact about black hole event horizons with positive surface gravity. For details the reader may consult \cite{Mihalisnotes}.

Let us investigate why this sign allows us to derive good estimates close to the horizon. The first step is to apply the identity (\ref{vecid}) with the vectorfield multiplier $N$ to the equation (\ref{commuteY}):
\begin{eqnarray} \label{mainidYp}
 \int_{\Sigma_\tau} J^N_\mu \left(\hat{Y}\psi\right)n^\mu_{\Sigma} +  \int_{\mathcal{H}^+\left(0,\tau\right)} J^N_\mu\left(\hat{Y}\psi\right) n^\mu_{\mathcal{H}} + \int_{\mathcal{R}\left(0,\tau\right) \cap \{r \leq r_0\}} \mathbf{K}^N\left(\hat{Y}\psi\right) \nonumber \\ = -\int_{\mathcal{R}\left(0,\tau\right) \cap \{r_0 \leq r \leq r_1\}} \mathbf{K}^N\left(\hat{Y}\psi\right) + \int_{\Sigma_0} J^N_\mu\left(\hat{Y}\psi\right) n^\mu_{\Sigma_0} \nonumber \\
+ \int_{\mathcal{R}\left(0,\tau\right) \cap \{r_{hoz} \leq r \leq r_0\}} \mathcal{E}^N\left(\hat{Y}\psi\right) + \int_{\mathcal{R}\left(0,\tau\right) \cap \{r > r_0\}} \mathcal{E}^N\left(\hat{Y}\psi\right)
\end{eqnarray}
where
\begin{eqnarray} \label{ENterms}
\mathcal{E}^N\left(\hat{Y}\psi\right) = -2e \frac{\Delta_-^\prime}{\Sigma} \left(\hat{Y}\hat{Y} \left(\psi\right)\right)^2  - \frac{2\Delta_-^\prime}{\Sigma} \left(\left(N - e\hat{Y}\right) \hat{Y}\psi\right) \left(\hat{Y}\hat{Y} \left(\psi\right)\right) \nonumber \\
+ \frac{4r}{\Sigma} \left(N\hat{Y}\psi\right)\left(\hat{Y}T\psi\right)  \nonumber \\
- \lambda_1 \left(\partial_r \psi\right)\left(N\hat{Y}\psi\right) - \lambda_2 \left(\partial_t \psi\right)\left(N\hat{Y}\psi\right) - \frac{\alpha}{l^2} \left(N\hat{Y}\psi\right) \left(Y\left(\psi\right)\right) \, .
\end{eqnarray} 
Let us start with the horizon term in (\ref{mainidYp}). As previously (cf.~(\ref{hozterm})) this flux has a good sign except for the lowest order term when $0<\alpha<\frac{9}{4}$.
%
%
 %
 %
This latter term is estimated as previously (cf.~(\ref{hoz2})) borrowing a bit from the good $\mathbf{K}^N$ term:
\begin{eqnarray}
\int_{\mathcal{H}^+\left(0,\tau\right)} \frac{e \gamma M}{2r k_0}\frac{\alpha}{l^2} \left(\hat{Y}\psi\right)^2\ \nonumber \\ \leq \epsilon \mathbf{K}^N\left(\hat{Y}\psi\right) + B \frac{1}{\epsilon} \int_0^\tau dt \int_{\Sigma_\tau} J^N_{\mu} \left(\psi\right) n^\mu \leq \epsilon \mathbf{K}^N \left(\hat{Y}\psi\right) + B D \tau \, ,
\end{eqnarray}
with
\begin{equation} \label{Ddef}
 D = \int_{\Sigma_0} \left(J^N_\mu \left(\psi\right) n^\mu_{\Sigma_0} + J^N_\mu \left(K \psi\right)n^\mu_{\Sigma_0}\right) \, .
\end{equation}
We turn to the $\mathcal{E}^N$-terms in (\ref{mainidYp}). Note that $N-e\hat{Y}=K$ on the horizon. Hence in $r\leq r_0$ (where weights in $r$ don't matter) we have
\begin{eqnarray}
 - \frac{2\Delta_-^\prime}{\Sigma} \left(\left(N - e\hat{Y}\right) \hat{Y}\psi\right) \left(\hat{Y}\hat{Y} \left(\psi\right)\right) \leq B \Big| \left( K+ \left(\left(1-\mu \right) \hat{Y} \right) \hat{Y}\psi\right) \left(\hat{Y}\hat{Y} \left(\psi\right)\right) \Big| \nonumber \\
\leq \frac{1}{\epsilon} \left(\hat{Y} K\psi\right)^2 + \epsilon \left(\hat{Y}\hat{Y} \left(\psi\right)\right)^2
\end{eqnarray}
by choosing $r_0$ sufficiently close to the horizon to exploit the $\left(1-\mu\right)$-term as a smallness factor. We are going to choose $\epsilon$ small enough to borrow from the good first term in (\ref{ENterms}). 
Similarly,
\begin{eqnarray}
\frac{4r}{\Sigma} \left(N\hat{Y}\psi\right)\left(\hat{Y}T\psi\right) \leq \epsilon \left(\hat{Y}\hat{Y} \left(\psi\right)\right)^2 + B \left(\hat{Y} K\psi\right)^2 + B |\lambda| \cdot \left(\partial_\phi \left(\hat{Y}\left(\psi\right)\right)\right)^2 .
\end{eqnarray}
Using several integrations by parts and the elliptic estimate (\ref{elliptic2}) we derive the following estimate for the last term:
\begin{eqnarray}
B |\lambda| \int_{\mathcal{R}\left(0,\tau\right)} \left(\partial_\phi \left(\hat{Y}\left(\psi\right)\right)\right)^2 \leq \epsilon \int_{\mathcal{R}\left(0,\tau\right) \cap \{r \leq r_0\}} \mathbf{K}^N\left(\hat{Y}\psi\right) + BD\left(\tau + 1\right) \nonumber \\ + \frac{1}{8} \int_{\mathcal{H}\left(0,\tau\right)} J^N_\mu \left(\hat{Y}\psi\right))n^\mu_{\mathcal{H}^+} + \frac{1}{8} \int_{\Sigma_\tau} J^N_\mu \left(\hat{Y}\psi\right)n^\mu_{\Sigma} +  \frac{1}{8} \int_{\Sigma_0} J^N_\mu \left(\hat{Y}\psi\right)n^\mu_{\Sigma} \, .
\end{eqnarray}
Let us denote the three boundary-terms in the last line collectively by $\mathcal{P}_1$. 

The other terms of $\mathcal{E}^N$ involve lower order terms and are estimated via Cauchy's inequality putting a small weight on the $\hat{Y}\hat{Y}$-term so that one can borrow again from the good term. Hence finally
\begin{eqnarray}
\int_{\mathcal{R}\left(0,\tau\right) \cap \{r \leq r_0\}} \mathcal{E}^N\left(\hat{Y}\psi\right) \leq \epsilon \int_{\mathcal{R}\left(0,\tau\right) \cap \{r \leq r_0\}} \mathbf{K}^N\left(\hat{Y}\psi\right) \nonumber \\ + \int_{\mathcal{R}\left(0,\tau\right) \cap \{r \leq r_0\}} \Bigg(\frac{1}{\epsilon} \left(\hat{Y}K\psi\right)^2 + \frac{1}{\epsilon} \mathbf{K}^N\left(\psi\right) \Bigg) + B_\epsilon D \left( \tau + 1\right) +\mathcal{P}_1 \nonumber \\ \leq \epsilon \int_{\mathcal{R}\left(0,\tau\right) \cap \{r \leq r_0\}} \mathbf{K}^N\left(\hat{Y}\psi\right) + B D \left( \tau+1\right) + \mathcal{P}_1\, .
\end{eqnarray}
The penultimate term in the second line deals in particular with the first order terms (which have the wrong sign if $\alpha >0$) arising in the $\mathbf{K}^N$ terms. For $r \geq r_0$ we only have to be careful with the weights in $r$:
\begin{eqnarray}
\int_{\mathcal{R}\left(0,\tau\right) \cap \{r \geq r_0\}} \mathcal{E}^N\left(\hat{Y}\psi\right) \leq \int dt \int_{\Sigma_\tau} \frac{\mathcal{E}^N\left(\hat{Y}\psi\right)}{\sqrt{-g^{t^\star t^\star}}} \, .
\end{eqnarray}
In the $AdS$ case $g^{t^\star t^\star} \sim \frac{1}{r^2}$ so in comparison with the asymptotically flat case (where this quantity approaches $1$ at infinity) one loses a power of $r$. However, the decay of $\mathcal{E}^N\left(\hat{Y}\psi\right)$ in $r$ is easily seen to be strong enough to have
\begin{eqnarray}
\int_{\mathcal{R}\left(0,\tau\right) \cap \{r \geq r_0\}} \mathcal{E}^N\left(\hat{Y}\psi\right) \leq \int_0^\tau dt \int_{\Sigma_\tau} \frac{\mathcal{E}^N\left(\hat{Y}\psi\right)}{\sqrt{-g^{t^\star t^\star}}} \nonumber \\ \leq \int_0^\tau dt \int_{\Sigma_\tau} \left(J^N_\mu \left(\psi\right) n^\mu_{\Sigma} + J^N_\mu \left(K \psi\right)\right) \leq B D \tau \, ,
\end{eqnarray}
using the elliptic estimate (\ref{elliptic1}) above. For $\mathbf{K}^N$ we have the analogous estimates to Lemma \ref{Kbulk}. In fact the lowest (first) order terms of the wrong sign can now be simply controlled by 
adding $\int_0^\tau dt \int_{\Sigma_\tau} J^N_{\mu} \left(\psi\right) n^\mu \leq B D \tau$ to the right hand side -- a Hardy inequality is no longer necessary. 

Putting these estimates together the identity (\ref{mainidYp}) turns into the estimate
\begin{eqnarray} 
 \int_{\Sigma_\tau} J^N_\mu \left(\hat{Y}\psi\right)n^\mu_{\Sigma} + b \int_0^\tau d\tau \int_{\Sigma_\tau} J^N_\mu \left(\hat{Y}\psi\right)n^\mu_{\Sigma} \nonumber \\ \leq B D \left( \tau + 1\right) + 2\int_{\Sigma_0} J^N_\mu\left(\hat{Y}\psi\right) n^\mu_{\Sigma_0} \, .
\end{eqnarray}
This is the analogue of (\ref{basic}) and hence
\begin{eqnarray} 
 \int_{\Sigma_\tau} J^N_\mu \left(\hat{Y}\psi\right)n^\mu_{\Sigma} \leq B \left(D + \int_{\Sigma_0} J^N_\mu \left(\hat{Y}\psi\right)n^\mu_{\Sigma_0}\right)
\end{eqnarray}
follows as previously. We have finally obtained control over the $rr$ derivative at the horizon. We now show how to control the second derivatives on the topological two-spheres close to the horizon defined by constant $(t^\star,r)$. For this we employ a second elliptic estimate. Write the wave equation as
\begin{eqnarray} \label{sphest}
\frac{1}{\Sigma} \left[\frac{1}{\sin \theta} \partial_\theta \left(\sin \theta \Delta_\theta \partial_\theta \psi\right) + \Sigma \left(g^{\phi \phi} - 2\lambda g^{t^\star\phi} + \lambda^2 g^{t^\star t^\star}\right) \partial^2_\phi \psi \right] \nonumber \\
= -g^{t^\star t^\star} \partial_t \left(K \psi\right) -2 g^{t^\star r} \partial_r \left(K \psi\right) + \left(g^{t^\star t^\star} \lambda - 2g^{t^\star \phi}\right) \partial_\phi \left(K \psi\right) \nonumber \\ - g^{rr} \partial_r \partial_r \psi + \xi \left(r-r_{hoz}\right) \partial_r \partial_\phi \psi + \textrm{lower order terms}
\end{eqnarray}
where $\xi$ is some bounded function. Note also the degenerating weight of $g^{rr}$ on the horizon. From (\ref{sphest}) we derive
\begin{equation}
 \int_{S_r} |\slashed{\nabla}^2 \psi|^2 \leq B \int_{S_r} \left(J^N_\mu \left(\psi\right) n^\mu_{\Sigma} + J^N_\mu \left(K\psi\right)n^\mu_{\Sigma} + J^N_\mu \left(\hat{Y}\psi\right)n^\mu_{\Sigma} \right)
\end{equation}
on a sphere of radius $r$. Close to the horizon we obtain 
\begin{equation} \label{elliptic2}
\int_{\Sigma_\tau \cap \{r\leq r_0 \}}  |\slashed{\nabla}^2 \psi|^2  \leq B \int_{\Sigma_\tau \cap \{r\leq r_0 \}} \left(J^N_\mu \left(\psi\right) n^\mu_{\Sigma} + J^N_\mu \left(K \psi\right) + \epsilon \cdot J^N_\mu \left(\hat{Y} \psi\right)  \right) \, ,
\end{equation}
with $\epsilon$ arising from choosing $r_0$ small enough to exploit 
the degenerating weights in (\ref{sphest}). Together with the previous elliptic estimate finally
\begin{eqnarray} \label{pen}
\int_{\Sigma} \left[\frac{1}{\sqrt{-g^{t^\star t^\star}}} h^{ij} h^{kl} \left(\nabla_i \nabla_k \psi \right) \left(\nabla_j \nabla_l \psi \right) + J^N\left(K\psi\right) n^\mu_{\Sigma} + J^N\left(\psi\right) n^\mu_{\Sigma} \right] \nonumber \\ \leq B \int_{\Sigma_\tau} \left(J^N_\mu \left(\psi\right) n^\mu_{\Sigma} + J^N_\mu \left( K\psi\right)  n^\mu_{\Sigma}+ J^N_\mu \left(\hat{Y} \psi\right)  n^\mu_{\Sigma} \right) \nonumber \\ \leq B \int_{\Sigma_0} \left(J^N_\mu \left(\psi\right) n^\mu_{\Sigma} + J^N_\mu \left(K\psi\right) n^\mu_{\Sigma_0} + J^N_\mu \left(\hat{Y}  \psi \right)  n^\mu_{\Sigma_0} \right) \, .
\end{eqnarray}
To write the estimate in a more geometric form we observe
\begin{equation}
\int_{\Sigma} \frac{1}{\sqrt{-g^{t^\star t^\star}}} \left[h^{ij} \nabla_i  \psi \nabla_j \psi \right] \leq B \int_{\Sigma} J^N_\mu\left(\psi\right) n^\mu_{\Sigma}
\end{equation}
and, from (\ref{KerrHardy}),
\begin{equation}
\int_{\Sigma} \frac{1}{\sqrt{-g^{t^\star t^\star}}} \psi^2 \leq B \int_{r_{hoz}}^\infty \int_{S^2} r^2 \psi^2 dr d\omega \leq B  \int_{\Sigma} J^N_\mu \left(\psi\right) n^\mu_{\Sigma} \, .
\end{equation}
The factor of $\frac{1}{\sqrt{-g^{t^\star t^\star}}} \sim r$ arises because $\sqrt{h} \sim r$ for the hyperbolic metric (instead of $r^2$ in the asymptotically flat case). In the other direction we have
\begin{eqnarray}
 \int_{\Sigma} J^N_\mu \left(\hat{Y}  \psi \right)  n^\mu_{\Sigma} \leq \nonumber \\
 B  \int_{\Sigma}\left(J^N\left(K\psi\right) n^\mu_{\Sigma} + J^N\left(\psi\right) n^\mu_{\Sigma} + \frac{1}{\sqrt{-g^{t^\star t^\star}}} h^{ij} h^{kl} \left(\nabla_i \nabla_k \psi \right) \left(\nabla_j \nabla_l \psi \right) \right) \, ,
\end{eqnarray}
which is seen by direct computation taking care of the weights of $r$. Hence the inequality (\ref{pen}) becomes
\begin{eqnarray}
\Big| \psi \Big|_{H^2_w\left(\Sigma\right)} +  \int_{\Sigma}\left(J^N\left(K\psi\right) n^\mu_{\Sigma} + J^N\left(\psi\right) n^\mu_{\Sigma}\right) \nonumber \\ \leq B \Bigg(\Big| \psi \Big|_{H^2_w\left(\Sigma_0\right)} +  \int_{\Sigma_0}\left(J^N\left(K\psi\right) n^\mu_{\Sigma_0} + J^N\left(\psi\right) n^\mu_{\Sigma_0}\right) \Bigg) 
\end{eqnarray}
with $H^2_w$ denoting the $\frac{1}{\sqrt{-g^{t^\star t^\star}}}$-weighted $H^2$ norm of $\Sigma$. Finally, we define
\begin{eqnarray}
\Big|n_\Sigma \psi\Big|_{H^1_\perp} := \int_{\Sigma} \frac{1}{\sqrt{-g^{t^\star t^\star}}} \left[r^2 \left(\nabla^{{\Sigma}} \cdot n_\Sigma \psi\right)^2 + \left(n_\Sigma\psi\right)^2\right] \nonumber \\ \leq B \Big( \Big| \psi \Big|_{H^2_w\left(\Sigma\right)} +   \int_{\Sigma}\left(J^N\left(K\psi\right) n^\mu_{\Sigma} + J^N\left(\psi\right) n^\mu_{\Sigma}\right) \Big)
\end{eqnarray}
and state the boundedness theorem for the massive wave equation on Kerr-AdS: 

\begin{theorem} \label{Kerrb}
Fix a cosmological constant $\Lambda = -\frac{3}{l^2}$, a mass $M>0$ and some $\alpha < \frac{9}{4}$. 
There exists an $a_{max}>0$, depending only on $\Lambda, M$ and $\alpha$, such that the following statement is true for all $a$ with $|a|<a_{max}$.

Let $\Sigma_0=\Sigma_{\tau_0}$ be a slice of constant $t^\star=\tau_0$ 
in $\mathcal{D}=\overline{J^+\left(\mathcal{I}\right) \cap J^-\left(\mathcal{I}\right)}$ 
of the Kerr-anti de Sitter spacetime $\left(\mathcal{M},g_{M,a,\Lambda}\right)$. 
Let $\psi$ be a solution to (\ref{wcf}) of class $C^{2}_{dec}$. If 
\begin{equation}
 \int_{\Sigma_0} J_\mu^K \left(\psi\right) n^\mu_{\Sigma_0} + \int_{\Sigma_0} J_\mu^K \left(K \psi\right) n^\mu_{\Sigma_0}< \infty
\end{equation}
then
\begin{eqnarray}
\Big| \psi \Big|_{H^2_w\left(\Sigma\right)} + \Big| n_\Sigma \psi \Big|_{H^1_{\perp}\left(\Sigma\right)} 
 + \sum_{m=0}^1 \int_{\Sigma} J_\mu^K \left(K^m \psi\right) n^\mu_{\Sigma} \nonumber \\ 
\leq C \Bigg(\Big| \tilde{\psi} \Big|_{H^2_w\left(\Sigma_0\right)} + \Big| n_\Sigma \tilde{\psi} \Big|_{H^1_{\perp}\left(\Sigma_0\right)} + \sum_{m=0}^1 \int_{\Sigma_0} J_\mu^K \left(K^m \psi\right) n^\mu_{\Sigma_0}\Bigg)
\end{eqnarray}
holds on $J^+\left(\Sigma_0\right) \cap \mathcal{D}$ for a uniform constant $C$ depending only on the parameters $M$, $a$, $l$ and $\alpha$. Here $\Sigma$ denotes any constant $t^\star$-slice to the future of $\Sigma_0$.
\end{theorem}

We remark that in contrast to the statement in the asymptotically flat case \cite{Mihalisnotes} we need the $K$ boundary term in this estimate. This is because the weighted $L^2$ norm of the second time-derivative that one obtains from the current $J^K$ is stronger than what can be derived from the wave equation in combination with the boundedness of $|\psi |_{H^2_w\left(\Sigma\right)}$ and $| n_\Sigma \psi|_{H^1_{\perp}\left(\Sigma\right)}$ alone. 

As mentioned previously, if $\alpha \leq 0$ then $a_{max} = \frac{r_{hoz}^2}{l}$ 
since all that was needed in the proof is that the vectorfield $K$ is timelike on the exterior (cf.~\cite{HawkingReall}). In general, the restriction on $a_{max}$ depends on $\alpha$ and becomes 
tighter as $\alpha$ approaches the Breitenlohner Freedman bound,  $\alpha \rightarrow \frac{9}{4}$. However, there 
is a uniform lower bound on $a_{max}$ (i.e.~$a_{max} \geq a_{uniform} > 0$ for \emph{any} $\alpha<\frac{9}{4}$) which can be worked out explicitly from the estimates of section 5.1. 

The Sobolev embedding theorem for asymptotically hyperbolic space yields
\begin{corollary}
We have the pointwise bound
\begin{eqnarray}
|\psi| \leq C \Bigg(\Big| \tilde{\psi} \Big|_{H^2_w\left(\Sigma_0\right)} + \Big| n_\Sigma \tilde{\psi} \Big|_{H^1_{\perp}\left(\Sigma_0\right)} + \sum_{m=0}^1 \int_{\Sigma_0} J_\mu^K \left(K^m \psi\right) n^\mu_{\Sigma_0}\Bigg)
\end{eqnarray}
on $J^+\left(\Sigma_0\right) \cap \mathcal{D}$.
\end{corollary}

To prove the corollary we rely on the following general Sobolev embedding theorem (cf. Theorem 3.4 in \cite{Hebey})
\begin{theorem}
Let $(N,h)$ be a smooth complete Riemannian $3$-manifold with Ricci curvature bounded from below and positive injectivity radius and $u \in H^2(N)$ a function on $N$. Then
\begin{equation} 
\sup_{N} |u|^2 \leq B \sum_{j=0}^2 \int_{N} |\nabla^j u|^2 dv\left(h\right) \, . 
\end{equation}
\end{theorem}
{\bf Remark: } Our $\Sigma_\tau$ is only complete with respect to the asymptotically hyperbolic end but it is straightforward to incorporate the boundary at $r=r_{hoz}>0$.

\section{Final Comments}
As mentioned in the abstract of the paper, the result does not make use of the separability properties of (\ref{study}) with respect to the Kerr background. In fact it does not make use of the axisymmetry either! All that was needed was the causal Killing vectorfield $K$ on the black hole exterior. In view of this fact, Theorem \ref{Kerrb} can be stated in the following generalized setting: \emph{Fix the Killing vectorfield $K=T+\lambda \Phi$ of a slowly rotating Kerr-AdS background as in section \ref{KerrAdS}. Perturb the metric such that it stays $C^1$-close to the Kerr-AdS metric and such that $K$ remains both Killing and null on the horizon.\footnote{The $C^1$ regularity is necessary because the surface gravity, whose positivity was essential for the argument, is $C^1$ in the metric.} Then  Theorem \ref{Kerrb} remains true for such spacetimes.} The main motivation for generalizations of this type are non-linear situations, in which the metric is not known explicitly a-priori but is itself dynamical.  In view of this one should use as less quantitative assumptions on the metric as possible to obtain bounds on the fields. Compare \cite{Mihalisnotes} for a more detailed discussion. As a further generalization one may assume only an approximate causal Killing field, i.e.~a vectorfield whose deformation tensor decays sufficiently fast in $t$. Treating the latter as a decaying error term in the estimates one can prove boundedness of (\ref{study}) for all spacetimes approaching a spacetime that is $C^1$-close in the sense above\footnote{i.e.~in particular admitting a timelike Killing field $K$ on the black hole exterior which becomes null on the horizon} to a slowly rotating Kerr-AdS solution. 

The question whether $\psi$ decays in time and if so, at what rate remains open. 

\section{Acknowledgments}
I would like to thank Mihalis Dafermos and Igor Rodnianski for stimulating discussions and useful comments. I am also grateful to an anonymous referee for his careful reading of the manuscript and many insightful remarks and suggestions.

\appendix
\section{Radial Decay} \label{weighted}
In this appendix we outline how -- assuming the boundary conditions (\ref{rdecI}) -- one can establish boundedness of appropriate higher weighted Sobolev-norms. The idea is the following. Given a solution $\psi$ of class $C^3_{dec}$ let $T^k\left(\psi\right)$ denote the application of $k$ times the vectorfield $T$ to $\psi$. We know that the $T$-energy associated with $T^k\left(\psi\right)$ is conserved for $k=0,1,2$. If we now revisit the wave equation to do elliptic estimates, this will introduce natural weighted Sobolev norms whose $r$-weights depend on $0<\alpha<\frac{9}{4}$. These optimized norms are expected to play a crucial role for the local existence theorem and are hence presented here.

Recall from (\ref{wofw}) that (for Kerr-AdS) one may write the wave equation in the form
\begin{eqnarray} \label{ap1}
\frac{1}{\sqrt{g}}\partial_i \left(g^{ij} \sqrt{g} \partial_j \psi \right) + \frac{\alpha}{l^2}\psi \nonumber \\ = -g^{t^\star t^\star}\left(\partial_{t^\star} \partial_{t^\star} \psi \right) - 2g^{t^\star i} \partial_{t^\star} \partial_i \psi - \frac{1}{\sqrt{g}} \partial_r \left(g^{t^\star r} \sqrt{g}\right) \left(\partial_{t^\star} \psi\right) \, .
\end{eqnarray}
The weighted $L^2$ norm of the right hand side decays very fast in view of the boundedness of $\int_\Sigma J^K_\mu\left(T^k\psi\right)n^\mu_{\Sigma} + J^K_\mu\left(\psi\right) n^\mu_{\Sigma}$. In particular, denoting the right hand side by $f$ we have
\begin{equation} \label{l2w}
\int_{r_{hoz}}^\infty \int_{S^2}dr d\omega r^4 f^2  < B \left(\int_\Sigma J^K_\mu\left(T\psi\right) n^\mu_\Sigma + J^K_\mu\left(\psi\right) n^\mu_\Sigma\right) \, .
\end{equation}
Let now $\chi=\chi\left(r\right)$ be a cut-off function which is equal to $1$ for $r>R$  for some fixed $R>r_{hoz}$ and is zero close to the horizon. Set $\sigma=\frac{2n}{n+3}$ for some $n<\sqrt{9-4\alpha}$ and multiply (\ref{ap1}) by 
\begin{eqnarray}
 \left(\frac{1}{\sqrt{g}}\partial_i \left(g^{ij} \sqrt{g} \chi \partial_j \psi \right) -  \chi \sigma \frac{\alpha}{l^2}\psi\right)\sqrt{g} r^n \nonumber \\ =  \chi \left(f + \left(1-\sigma\right))\frac{\alpha}{l^2} \psi\right)\sqrt{g} r^n + \chi_{,r} g^{rj} \partial_j \phi \sqrt{g} r^n\nonumber \, .
\end{eqnarray}
Integrating over the slice $\Sigma$, we observe (in view of (\ref{l2w})) that we can bound the right hand side as long as $n \leq 2$ (provided we can borrow an $\epsilon$ of the term $\int r^{2+n} \psi^2 dr d\omega$ from the left). 
The radial derivatives\footnote{these derivatives are the problematic ones} on the left hand side can after several integrations by parts be estimated by 
\begin{eqnarray} \label{helpit}
\int_{R}^\infty \int_{S^2} dr d\omega r^{6+n} \left(\partial_r \partial_r \psi\right)^2  \nonumber \\ +\int_{R}^\infty \int_{S^2} dr d\omega \left(16-4\left(5+n\right)-\alpha + \alpha \sigma\right) r^{4+n} \left(\partial_r \psi\right)^2 \nonumber \\ \int_{R}^\infty \int_{S^2} dr d\omega \alpha \left(-\sigma \alpha + \frac{1-\sigma}{2} n \left(n+3\right) \right)r^{2+n} \psi^2 \nonumber \\ \leq B \left(\int_\Sigma J^K_\mu\left(T\psi\right) n^\mu_\Sigma + \int_\Sigma J^K_\mu \left(\psi\right) n^\mu_\Sigma\right) \, .
\end{eqnarray}
The boundary terms arising in the computations all vanish: on the left because of the cut-off function, and on the right  in view of (\ref{rdecI}) (the restriction $n<\sqrt{9-4\alpha}$ is imposed in order to make these boundary terms vanish). Note also that the error-terms introduced by the derivatives of the cut-off function are well in the interior (bounded $r$) and hence unproblematic. The coefficient of the zeroth order term in (\ref{helpit}) is positive if $\frac{2n}{n+3}=\sigma<\frac{n\left(n+3\right)}{2\alpha + n\left(n+3\right)}$, which an easy calculation reveals to be true for $n<\sqrt{9-4\alpha}$. (In particular we have room to borrow the aforementioned $\epsilon$ of this term for the right hand side.) The second term on the left of (\ref{helpit}) is negative. How large can we allow $n$ to be to still absorb this term by the first term using a Hardy inequality? As one easily checks, the Hardy condition
\begin{equation}
 \left(4+4n+\alpha\left(1-\sigma\right)\right)\frac{4}{\left(n+5\right)^2} < 1
\end{equation}
is satisfied for
\begin{equation}
n < \sqrt{9-4\alpha} \, .
\end{equation}
We conclude that away from the horizon
\begin{eqnarray} \label{elliptic1b}
\int_{\Sigma \cap \{r \geq R \}} \frac{r^n}{\sqrt{-g^{t^\star t^\star}}} h^{ij} h^{kl} \left(\nabla_i \nabla_k \psi \right) \left(\nabla_j \nabla_l \psi \right) \nonumber \\ 
\leq C \left(R, n\right) \int_{ \Sigma \cap \{r \geq r_{hoz} \}} \left(J^K_{\mu}\left(\psi\right) n^\mu_{\Sigma} + J^K_{\mu}\left(T \psi\right) n^\mu_{\Sigma}\right) 
\end{eqnarray}
for any $n<\min\left(2,\sqrt{9-4\alpha}\right)$. In case that $\sqrt{9-4\alpha}>2$ one simply repeats the argument after another commutation with $T$. In particular, (\ref{l2w}) holds with a weight $r^6$ replacing $r^4$ on the left, if one adds the term $\int_\Sigma J^K_\mu\left(T T\psi\right) n^\mu_\Sigma$ on the right hand side.\footnote{This in turn is a consequence of the Hardy inequality $\int r^2 \left(\partial_t \partial_t \psi\right)^2 dr \leq \frac{4}{9} \int r^4 \left(\partial_r \partial_t \partial_t \psi\right)^2 dr$.} From this one derives estimate (\ref{elliptic1b}) with any weight $n<\sqrt{9-4\alpha}$ adding $\int_\Sigma J^K_\mu\left(T T\psi\right) n^\mu_\Sigma$ on the right hand side. 

\bibliographystyle{hacm}
\bibliography{thesisrefs}
\end{document}

%% file: conformal2.pstex_t
\begin{picture}(0,0)%
\includegraphics{conformal2.pstex}%
\end{picture}%
\setlength{\unitlength}{1145sp}%
\begingroup\makeatletter\ifx\SetFigFont\undefined%
\gdef\SetFigFont#1#2#3#4#5{%
  \reset@font\fontsize{#1}{#2pt}%
  \fontfamily{#3}\fontseries{#4}\fontshape{#5}%
  \selectfont}%
\fi\endgroup%
\begin{picture}(3462,6699)(589,-6148)
\put(4051,-2311){\makebox(0,0)[lb]{\smash{{\SetFigFont{5}{6.0}{\rmdefault}{\mddefault}{\updefault}{\color[rgb]{0,0,0}$\mathcal{I}$}%
}}}}
\put(1651,-1261){\makebox(0,0)[lb]{\smash{{\SetFigFont{5}{6.0}{\rmdefault}{\mddefault}{\updefault}{\color[rgb]{0,0,0}$\mathcal{H}^+$}%
}}}}
\put(2101,-2611){\makebox(0,0)[lb]{\smash{{\SetFigFont{5}{6.0}{\rmdefault}{\mddefault}{\updefault}{\color[rgb]{0,0,0}$\Sigma_0$}%
}}}}
\put(2401,-1711){\makebox(0,0)[lb]{\smash{{\SetFigFont{5}{6.0}{\rmdefault}{\mddefault}{\updefault}{\color[rgb]{0,0,0}$\Sigma_\tau$}%
}}}}
\end{picture}%